\documentclass[sigconf]{acmart}

\AtBeginDocument{%
  }

\usepackage{balance}
\usepackage{verbatim}
\usepackage{subfigure}
\usepackage{enumitem}
\usepackage{diagbox}
\usepackage{booktabs}
\usepackage[ruled,vlined,linesnumbered,boxed,commentsnumbered]{algorithm2e}  
\usepackage{algpseudocode}
\usepackage{color}
\usepackage{url}
\usepackage{multirow}

\usepackage{lipsum}
\usepackage{tipa}
\usepackage{xcolor}
\usepackage{framed}
\usepackage{soul}

\usepackage{tikz}
\usepackage{pgfplots}
\newcommand*\circled[1]{\tikz[baseline=(char.base)]{
    \node[shape=circle,draw,inner sep=0.5pt] (char) {\small#1};}}

\newcommand{\del}[1]{{}}
\newcommand{\add}[1]{{{#1}}}
\newcommand{\change}[2]{{#2}}

\newcommand{\qingchao}[1]{{\color{teal}[Qingchao: #1]}} 
\newcommand{\jj}[1]{{\color{orange}[Junjie: #1]}}

\newcommand{\tool}{\textsc{OATest}}

\newcommand{\llmbaseline}{\textsc{LLMTest}}

\newcommand{\TotalBugsNum}{$56$} 
\newcommand{\TotalTVMBugsNum}{$40$}
\newcommand{\TotalORTBugsNum}{$16$}

\newcommand{\OPTBugsNum}{$33$}
\newcommand{\TVMOPTBugsNum}{$24$}
\newcommand{\ORTOPTBugsNum}{$9$}

\newcommand{\TotalBugsNumCC}{$42$}
\newcommand{\TotalTVMBugsNumCC}{$29$}
\newcommand{\TotalORTBugsNumCC}{$13$}

\newcommand{\ConfirmedBugsNum}{$42$}

\newcommand{\FixedBugsNum}{$24$}

\newcommand{\TotalConfirmedCrashBugs}{$36$}

\newcommand{\TotalConfirmedWrongBugs}{$6$}
\newcommand{\NNSmithBugsNum}{$5$}
\newcommand{\WhiteFoxBugsNum}{$3$}

\newcommand{\TVMDonorNum}{$942$}
\newcommand{\ORTDonorNum}{$2,116$}

\newcommand{\BranchcovImporve}{$60.20\%$}
\newcommand{\LinecovImporve}{$66.98\%$}

\newcommand{\pair}[1]{$\langle {#1} \rangle$}
\newcommand{\mycode}[1]{\texttt{#1}\xspace}

\newcommand{\eg}{\hbox{\emph{e.g.}}\xspace}
\newcommand{\ie}{\hbox{\emph{i.e.}}\xspace}

\newcommand{\viz}

\copyrightyear{2026}
\acmYear{2026}
\setcopyright{cc}
\setcctype{by}
\acmConference[ICSE '26]{2026 IEEE/ACM 48th International Conference on Software Engineering}{April 12--18, 2026}{Rio de Janeiro, Brazil}
\acmBooktitle{2026 IEEE/ACM 48th International Conference on Software Engineering (ICSE '26), April 12--18, 2026, Rio de Janeiro, Brazil}
\acmPrice{}
\acmDOI{10.1145/3744916.3773216}
\acmISBN{979-8-4007-2025-3/26/04}

\begin{document}

\title{Optimization-Aware Test Generation for Deep Learning Compilers}

\author{%
Qingchao Shen\mbox{\textsuperscript{1}},
Zan Wang\mbox{\textsuperscript{1}},
Haoyang Ma\mbox{\textsuperscript{2}},
Yongqiang Tian\mbox{\textsuperscript{3}},
Lili Huang\mbox{\textsuperscript{1}},
Zibo Xiao\mbox{\textsuperscript{1}}
\\Junjie Chen\mbox{\textsuperscript{1,*}},
Shing-Chi Cheung\mbox{\textsuperscript{2}}
}
\affiliation{
  \institution{
  \textsuperscript{1}School of Computer Software, Tianjin University \city{Tianjin} \country{China}\\
  Email: \{qingchao, wangzan, huangll, zibo.xiao, junjiechen\}@tju.edu.cn\\
  \textsuperscript{2}The Hong Kong University of Science and Technology\city{Hong Kong} \country{China}\\
  Email: haoyang.ma@connect.ust.hk, scc@cse.ust.hk\\
  \textsuperscript{3}Monash University \city{Melbourne} \country{Australia},
  Email: yongqiang.tian@monash.edu\\
  }
}
\thanks{\textsuperscript{*}Junjie Chen is the corresponding author.}










\renewcommand{\shortauthors}{Shen et al.}

\begin{abstract}
Deep Learning (DL) compilers have been widely utilized to optimize DL models for efficient deployment across various hardware. Due to their vital role in the DL ecosystem, ensuring their reliability is critical. Such model optimizations are often designed to match specific computational graph structures.
However, existing DL compiler fuzzing techniques do not generate tests for each optimization aware of its matched graph structures.
In this paper, we propose OATest, a novel technique that synthesizes optimization-aware tests by extracting patterns from documented tests and integrating them into diverse computational graph contexts.
To address the key technical challenge of synthesizing valid and optimization-aware computational graphs, OATest introduces two synthesis strategies: (1) reusing compatible inputs/outputs from existing nodes and (2) creating new nodes with compatible inputs/outputs, establishing effective connections between patterns and contexts.
OATest is evaluated on two popular DL compilers, TVM and ONNXRuntime, regarding the bugs revealed by crashes and inconsistencies.
The experimental results show that OATest significantly outperforms the state-of-the-art DL compiler fuzzing techniques by detecting more bugs and covering more optimization code in TVM and ONNXRuntime. Particularly, OATest uncovers 56 previously unknown bugs, 42/24 of which have been confirmed/fixed by developers. 
\end{abstract}


\keywords{Compiler Testing, Fuzzing, Test Synthesis, Deep Learning Compiler}


\renewcommand\footnotetextcopyrightpermission[1]{}

\maketitle
\setlength{\abovecaptionskip}{1.7mm}
\setlength{\belowcaptionskip}{-2mm}
\section{Introduction}
\label{sec:intro}
Deep Learning (DL) compilers (\eg, TVM~\cite{tvm}, ONNXRuntime~\cite{ort}, and TensorRT~\cite{TensorRT}) have become increasingly popular and play a fundamental role in optimizing DL models for efficient deployment on diverse hardware platforms.
However, DL compilers are also vulnerable to bugs, in which compiler optimization (the key component) is more error-prone than other components~\cite{shen2021comprehensive}.
Moreover, due to the fundamental role of DL compiler optimizations, their bugs could produce serious consequences~\cite{polyjuice}, 
causing compiled models to execute incorrectly (leading to unintended behavior)
, or even causing runtime failures on specific hardware (leading to service unavailability).
Therefore, it is imperative to ensure the quality of DL compiler optimizations.


In recent years, several DL compiler fuzzing techniques~\cite{NNSmith,ma2022hirfuzz,whitefox} have been proposed. 
The most widely-studied ones are grammar-based techniques, such as NNSmith~\cite{NNSmith} and HirGen~\cite{ma2022hirfuzz}.
They focus on generating valid tests, \ie, computational graphs (with nodes representing operators and edges signifying data dependencies between operators), based on grammar for fuzzing DL compilers. 
However, triggering an optimization tends to require specific patterns~\cite{whitefox}, \eg, only the structure of \mycode{Concat(PermuteDims(A), PermuteDims(B))}, which first rearranges the dimensions of tensors \mycode{A} and \mycode{B} using \mycode{PermuteDims} and then concatenates them into a single tensor, can be optimized by the \mycode{ReorderPermuteDimsAfterConcat} optimization (also called pass in the field of compilers) and thus may trigger potential bugs in this optimization.
Such grammar-based techniques just arbitrarily construct tests based on grammar without special attention on diverse optimizations, leading to limited effectiveness in detecting optimization bugs.
For example, NNSmith covers only 24.75\% of TVM's optimization code branches during 12-hours fuzzing (to be presented in Section~\ref{sec:result_coverage}).



Unlike grammar-based techniques, WhiteFox~\cite{whitefox} was introduced recently, which leverages Large Language Models (LLMs) to mine optimization patterns from compiler source code and generate tests based on them. 
While WhiteFox has successfully uncovered a certain number of optimization bugs,
it still has two main limitations.
First, WhiteFox relies on the capability of LLMs on code comprehension, 
but the compiler source code is much more complicated, 
limiting its effectiveness in extracting correct patterns and thus generating optimization-aware tests.
For example, the \mycode{DataflowInsertInPlaceCalls} pass requires an in-place operator, but the pattern extracted by WhiteFox (\ie, \mycode{t2 = t1 + t1}) is creating a new tensor (\ie, \mycode{t2}) rather than modifying \mycode{t1} in place, which does not align with this optimization.
Second, LLMs' hallucinations limit the effectiveness of generating valid tests, thereby limiting the overall fuzzing effectiveness. 
For instance, more than half of the tests generated by WhiteFox are invalid (as detailed in Section~\ref{sec:test_quality}) and have to be rejected before any optimizations are performed.
This motivates more research efforts to bridge the gap in designing effective test generation techniques for DL compiler optimizations.
To improve the effectiveness in detecting DL compiler optimization bugs, the generated tests should be as optimization-aware and diverse as possible.
To enhance the possibility of triggering DL compiler optimizations, we choose a more straightforward source, \ie, documented tests for each optimization, to learn optimization-aware patterns.
On the one hand, these tests are manually written by developers to test the corresponding optimizations and thus embody the specific optimization-aware patterns.
On the other hand, such patterns can be extracted through computational graph analysis according to the optimization characteristics, which is a more deterministic manner and thus avoids the non-determinism and inaccuracy incurred by LLMs in comprehending complicated compiler source code.
To increase test diversity, we synthesize tests by integrating a pattern corresponding to an optimization into different contexts.
Given the context-sensitive nature of DL compiler optimizations~\cite{tvm}, this integration may trigger different optimization behaviors, indicating the construction of diverse optimization scenarios to some degree.

Based on these insights, we propose our idea of computational graph synthesis by integrating the optimization-aware patterns learned from documented tests into diverse contexts for improving the effectiveness of fuzzing DL compiler optimizations.
Specifically, \tool{} contains three main steps.
First, \tool{} extracts optimization-aware patterns from the computational graphs gathered from documented tests via code instrumentation. 
Here, different optimizations work at different granularities (\ie, block - a self-contained sequence of operations, and subgraph - a connected portion of a computation graph)~\cite{dlc_survey}, and thus \tool{} extracts block-level or subgraph-level patterns accordingly in order to ensure the optimization awareness of the extracted patterns.
Second, \tool{} synthesizes tests by integrating extracted patterns into different contexts provided by a set of existing tests. 
Simply putting them together will lead to invalid tests (\ie, undefined inputs for the operators in patterns), \tool{} designs two strategies to fix such problems, which is also its key technical challenge.
To establish effective connections between patterns and contexts, one fixing strategy in \tool{} is to reuse the compatible inputs/outputs of the existing nodes in the context for creating edges to the nodes with undefined inputs in the pattern.
This is helpful to enhance the interaction between patterns and contexts for activating new optimization behaviors.
As it is hard to ensure the existence of compatible inputs/outputs in the context, another fixing strategy is complementary, which creates new nodes with compatible inputs/outputs and then connects them to the nodes with undefined inputs in the pattern.
Particularly, \tool{} enables iterative synthesis of patterns across diverse contexts, facilitating the combination of multiple patterns for fuzzing DL compiler optimizations.
Third, \tool{} incorporates two test oracles (\ie, crash and inconsistency) to determine whether synthesized tests reveal DL compiler bugs.

We evaluated \tool{} through an extensive study on two widely-used DL compilers, \ie, TVM~\cite{tvm} and ONNXRuntime~\cite{ort}.  
\tool{} extracted \TVMDonorNum{} and \ORTDonorNum{} 
patterns from documented tests in TVM and ONNXRuntime, respectively.
Here, we used the tests generated by NNSmith (a grammar-based test generation technique for DL compilers) as the context provider for computation graph synthesis by \tool{}, which are called seed graphs for ease of presentation.
Over a two-week fuzzing period, \tool{} uncovered a total of \TotalBugsNum{} previously unknown bugs, of which \ConfirmedBugsNum{}/\FixedBugsNum{} were confirmed/fixed by developers. 
The bugs detected by \tool{} are diverse, involving a wide range of optimizations and root causes.
We also compared \tool{} with two state-of-the-art DL compiler fuzzers (\ie, WhiteFox and NNSmith) \add{and a custom LLM-based technique we developed (\ie, \llmbaseline{})}.
Within a 12-hour fuzzing budget for repeating 5 times, \tool{} detected \TotalBugsNumCC{} unique bugs (including 30 optimization bugs), significantly outperforming NNSmith\change{ and}{,} WhiteFox\add{ and \llmbaseline{}}, which identified only 5 bugs (including 2 optimization bugs)\change{ and}{,} 4 bugs (including 3 optimization bugs)\add{ and 7 bugs (including 3 optimization bugs)}, respectively.
Moreover, \tool{} covered \change{61.15\%}{\BranchcovImporve{}} more branches and \change{59.56\%}{\LinecovImporve{}} more lines on average compared to the \change{two}{three} baselines regarding DL compiler optimization code, demonstrating the significant superiority of \tool{} over the state-of-the-art fuzzing techniques. 
Also, our evaluation demonstrates the efficiency of \tool{} and confirms the contribution of key components in \tool{}.

To sum up, this work makes the following major contributions:
\begin{itemize}[topsep=0pt,leftmargin=*]
    \item We introduce \textit{computational graph synthesis} to generate optimization-aware tests for DL compiler optimizations.
    
    \item We develop a novel test generation technique (\ie, \tool{}) guided by the idea, which extracts optimization-aware patterns from documented tests and integrates them with diverse contexts to synthesize optimization-aware computational graphs as tests.

    \item We conduct an extensive study to evaluate \tool{} based on two widely-used DL compilers (\ie, TVM and ONNXRuntime), demonstrating the superiority of \tool{} over two state-of-the-art DL compiler fuzzing techniques by detecting \TotalBugsNum{} previously unknown bugs with \ConfirmedBugsNum{}/\FixedBugsNum{} confirmed/fixed.
    
\end{itemize}

The source code of \tool{} and all experimental data have been released for replication and future use, which are accessible at: \textbf{\url{https://github.com/ShenQingchao/OATest}.}

\section{Preliminaries}  
\label{sec:background}

\subsection{DL Compilers}
\label{dl_compiler}



DL compilers process DL models to produce optimized, deployable code for diverse hardware. 
In general, DL compilers involve two main stages.
First, the \textit{model loading} stage is responsible for converting 
DL models built from various DL libraries (\eg, PyTorch~\cite{pytorch} and TensorFlow~\cite{tensorflow}) into
Intermediate Representations (IRs).
DL models and IRs are two distinct formats of computational graphs:
the former is designed for inference while the latter is designed for optimization.
Regardless of the formats, a computation graph is a directed acyclic graph, where nodes represent operators (\eg, Conv2D), edges represent data dependencies between operators, and tensors flow along edges as inputs and outputs of operators. That is, the variable on the edge is a tensor representing the input or output of an operator.
Then, the \textit{optimization} stage applies a sequence of optimizations (also called \textit{passes}) to the IR for transformation in order to enhance its performance.

Optimizations in DL compilers can be classified into two categories:
block-level optimization (\eg, algebraic simplification) and subgraph-level optimization (\eg, common subexpression elimination) to reduce redundancy and enhance efficiency~\cite{dlc_survey}.
The optimization can only be executed when the computational graph contains a specific structure that matches the predefined conditions in the optimization (\ie, pattern).
For example, when a computational graph contains a (\eg, \mycode{Add+ReLU}) structure that matches the pattern of \mycode{OpFusion} \textit{pass}, the \textit{pass} will fuse the two operators into one to form a more optimized version.
Finally, the optimized IR is compiled to generate deployable code for diverse hardware.

Unlike traditional compilers (\eg, LLVM~\cite{llvm} and JVM~\cite{jvm_jit}), DL compilers 
take as input computational graphs instead of imperative programs and execute various DL-specific optimizations (\eg, operator fusion).
Consequently, fuzzing techniques~\cite{javatailer,emi,csmith} for traditional compilers cannot be applied to DL compilers due to significant differences in input formats and test objectives~\cite{shen2021comprehensive,opera}.

\subsection{A Motivating Example}
\begin{figure}
    \centering
    \includegraphics[width=\linewidth]{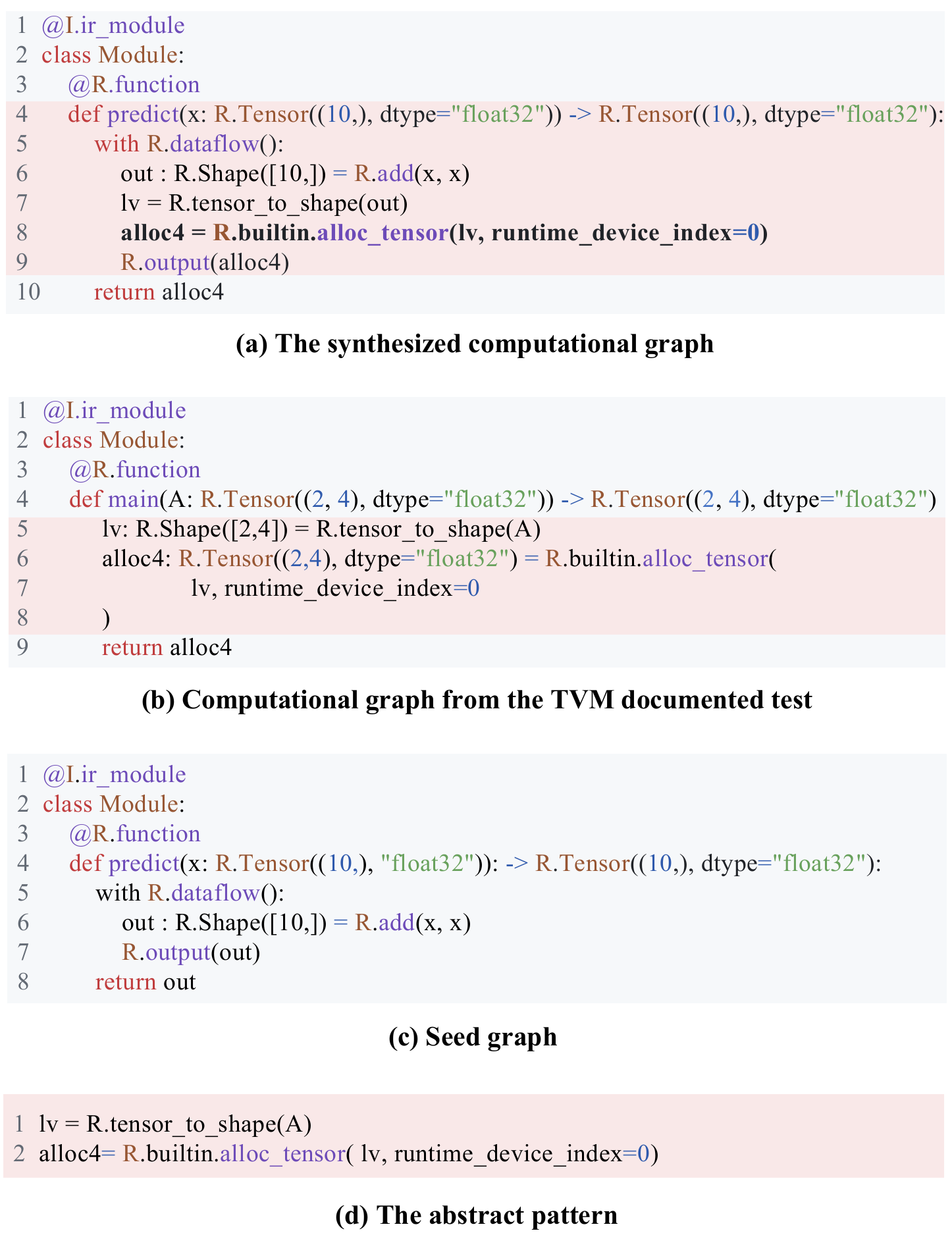}
    \caption{Motivation example}
    \label{fig:motivation_example}
\end{figure}

Figure~\ref{fig:motivation_example}(a) shows a computational graph generated by \tool{} that detects a previously unknown bug in TVM~\cite{issue_motivation}.
This bug in the \mycode{StaticPlanBlockMemory} \textit{pass} stems from the incorrect handling for the \mycode{DataflowBlockNode} instance (\ie, \mycode{R.builtin.alloc\_tensor}).
The \mycode{StaticPlanBlockMemory} \textit{pass} is a block-level optimization that enhances memory allocation by analyzing usage patterns and lifetimes for static planning and reuse.
However, when utilizing the \mycode{StaticPlanBlockMemory} \textit{pass} to optimize the synthesized graph, TVM crashed unexpectedly.
After reporting to developers, it has been fixed by introducing handling for the \mycode{DataflowBlockNode}, ensuring proper processing of blocks during memory planning~\cite{pr_motivation}.

This bug is difficult to trigger by existing techniques because it requires strict conditions, including 
(1) a specific pattern (\ie, a static computational graph with the \mycode{R.builtin.alloc\_tensor} operator) for triggering the optimization and 
(2) an extra condition (\ie, the pattern must be wrapped in a dataflow block -- the \mycode{R.dataflow()} construct in Line 5 of Figure~\ref{fig:motivation_example}(a))
to cover the buggy code.
Among 1,000 tests generated by NNSmith, none triggered this optimization, confirming the limited effectiveness of such a grammar-based technique.
Also, WhiteFox fails to extract complete triggering conditions and always produces invalid tests due to LLMs' hallucination, thereby missing this bug.

We noticed that the tests documented in TVM include the specific pattern but fail to reveal the bug because they lack the second context condition, resulting in the buggy code remaining uncovered.
A bug-revealing computational graph (\ie., Figure~\ref{fig:motivation_example}(a)) can be generated by synthesizing the optimization-aware pattern extracted from the documented test (\ie., Figure~\ref{fig:motivation_example}(b)) with a seed graph generated by NNSmith (\ie., Figure~\ref{fig:motivation_example}(c)).
Note that the seed graph also cannot trigger the bug due to lacking the pattern for triggering the \mycode{StaticPlanBlockMemory} \textit{pass}.
Specifically, we first extract the optimization-aware pattern (\ie, Lines 5-8 in Figure~\ref{fig:motivation_example}(b)) from the computational graph collected from the TVM documented test.
Then, we synthesize a new test by combining the pattern with a seed graph that contains the extra condition.
During the synthesis process, a technical challenge is to establish effective connections between the pattern and the context and thus generate a valid test.
Specifically, direct insertion would result in an undefined input for \mycode{R.tensor\_to\_shape} (\ie, the variable \mycode{A}), which has a shape constraint of \mycode{(2,4)}.
Moreover, we cannot find a compatible variable in the seed graph that satisfies the shape constraint for fixing the undefined inputs.
\add{
Our empirical evaluation revealed that direct test insertion achieved merely 5\% validity rate during a 12-hour fuzzing process.
}
To solve it, we abstract the extracted pattern into an abstracted one to loosen the constraints by eliminating all specific shape and type values in the inputs/outputs of nodes
(\ie., Figure~\ref{fig:motivation_example}(d)) and fix the undefined input (\eg, parameters of \mycode{A}) in the abstract pattern by reusing the input of the existing node (\eg, the \mycode{out} variable in Figure~\ref{fig:motivation_example}(c)) in the seed graph.
Note that if none of the variables in the context satisfy the connection constraints, we will create bridge nodes to link the undefined input in the pattern to the seed graph
, thereby forming a valid computational graph (will be introduced in Section~\ref{sec:auxiliary_adding}).
In this way, the bug-revealing computational graph is finally generated (as shown in Figure~\ref{fig:motivation_example}(a)).

Intuitively, mutation-based test generation that mutates the documented tests may also generate tests to fuzz DL compiler optimizations.
However, \change{optimizations in DL compilers focus on relatively coarse-grained blocks or subgraphs in computational graphs, making it difficult to generate diverse optimization-aware tests through minor mutations (\eg, mutating the parameter settings for operators).}{no mutation-based technique exists specifically for DL compiler testing. Traditional mutation (\eg, minor changes for the parameter settings of operators) is less effective here due to DL compilers’ coarse-grained blocks or subgraphs optimizations in computational graphs.}
This restricts their ability to cover new optimization behaviors, reducing their effectiveness in fuzzing DL compiler optimizations~\cite{opera}.
In summary, embedding optimization-aware patterns into diverse contexts can be more effective in generating optimization-aware tests.

\section{Approach}
\label{sec:methodology}

\begin{figure*}[]
    \centering
    \includegraphics[width=1\linewidth]{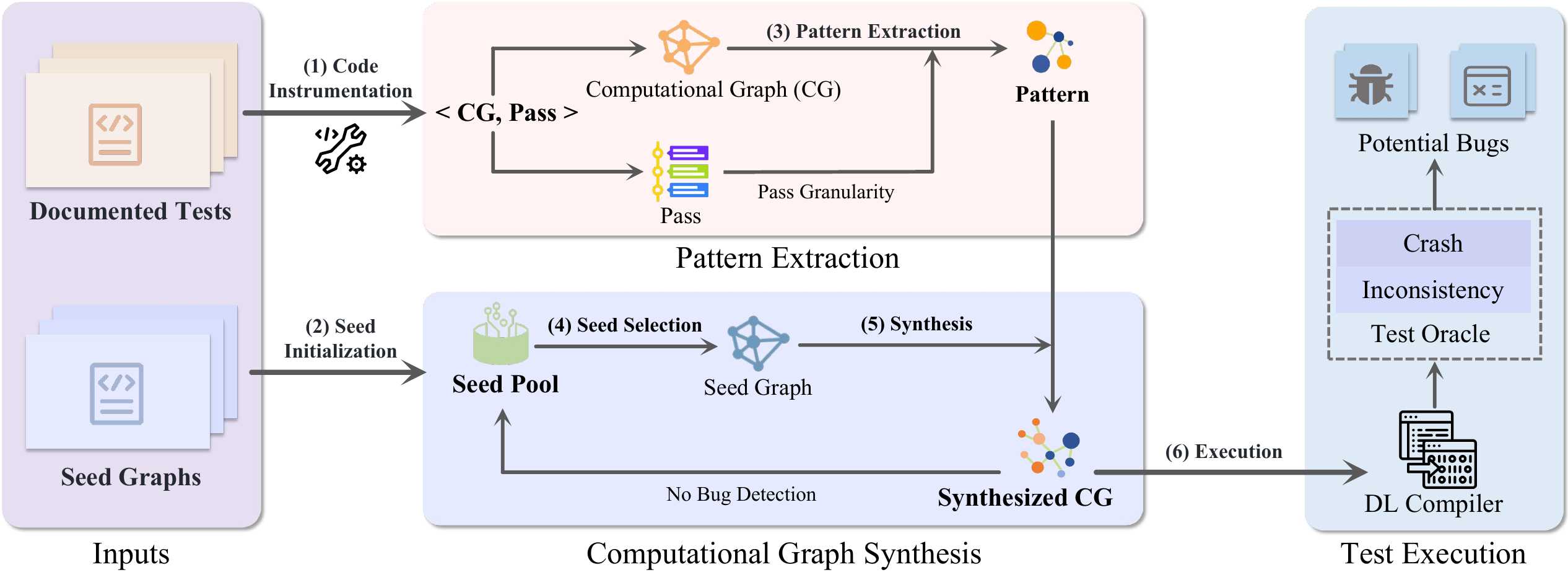}
    \caption{The Overview of \tool{}}
    \label{fig:overview}
\end{figure*}

In this section, we present the methodology of \tool{}, our novel DL compiler optimization fuzzing technique, which aims to synthesize computational graphs by integrating optimization-aware patterns obtained from documented tests into diverse contexts. 
As illustrated in Figure~\ref{fig:overview}, \tool{} comprises three primary steps.
First, \tool{} extracts optimization-aware patterns from computational graphs by instrumenting code in documented tests.
As optimizations in DL compilers work at different granularities (\ie, block and subgraph), 
patterns are identified by extracting blocks or subgraphs accordingly as needed to ensure optimization awareness (Section~\ref{sec:pattern}).
Second, \tool{} synthesizes a computational graph by incorporating an extracted pattern into a given
seed graph,
where the seed graph serves as context (Section~\ref{sec:synthesis}). 
In particular, \tool{} fixes the undefined inputs of operators in the pattern 
by either 
(1) utilizing the compatible inputs/outputs of existing nodes in the context or 
(2) creating nodes with compatible inputs/outputs as a bridge for connection.
Third, \tool{} incorporates two test oracles to run the test and determine whether a synthesized computational graph uncovers a DL compiler bug.
Bug-free computational graphs
generated by \tool{}
will be added to the seed pool as a new seed graph
for further synthesizing with pattern combination
(Section~\ref{sec:oracle}).

\subsection{Pattern Extraction}
\label{sec:pattern}
In documented tests of DL compilers, diverse optimization-aware computational graphs \add{(CGs)} were crafted to test each optimization's implementation.
Since different optimizations work at different granularities, extracting optimization-aware patterns from the documented tests involves two steps:
(1) identifying the computational graph and its corresponding optimization (denoted as \pair{CG, \textit{pass}} pair), and (2) extracting optimization-aware patterns from the computational graph based on the granularity of the optimization.

First, \tool{} collects the \pair{CG, \textit{pass}} pairs from documented tests. 
In DL compilers, documented tests only include computational graph generation statements,
instead of a concrete computational graph. 
These statements dynamically generate a computational graph in the memory during test execution.
Since static analysis is insufficient for capturing such runtime-generated computational graphs,
\tool{} utilizes code instrumentation to dynamically gather all \pair{CG, \textit{pass}} pairs during test execution in DL compilers. 
Specifically, to capture all \pair{CG, \textit{pass}} pairs in the documented tests, 
\tool{} hooks into the \textit{pass} call function in the base class of \textit{pass}, 
which is inherited by all specific \textit{passes}. 
This function allows access to both the \textit{pass} object and the computational graph being optimized.
Once the hooked API is called, 
the computational graph to be optimized and the invoked optimization will be collected simultaneously as a \pair{CG, \textit{pass}} pair, respectively.


For each collected \pair{CG, \textit{pass}} pair, \tool{} further extracts the optimization-aware patterns from the computational graph based on the granularity of the associated optimization.
Since each optimization is thoroughly described in the documentation of the corresponding DL compiler, we identify its granularity by analyzing the corresponding description and categorizing it as either block-level or subgraph-level.
Specifically, 
\tool{} begins by constructing a dependency graph that captures the relationships between nodes (\ie, operators) in the computational graph. It then traverses the graph to identify the corresponding patterns.
For subgraph-level optimization, \tool{} identifies distinct computational units as subgraph patterns by grouping operators that share input or output tensors. 
For block-level optimization, \tool{} further groups operations within the subgraph into block patterns based on shared intermediate tensors or execution order dependencies.
\add{A subgraph is a connected, semantically complete computational unit within a computational graph, with explicit input/output boundaries and executable isolation, while a block is a structured, contiguous region of a subgraph defining a localized computation.}
This method efficiently decomposes the computational graph into meaningful optimization-aware patterns prepared for subsequent computational graph synthesis.

\subsection{Computational Graph Synthesis}
\label{sec:synthesis}

During the computational graph synthesis process, \tool{} randomly selects a pattern and a seed graph, then inserts the pattern into the seed graph by randomly choosing the synthesis point. 
These random selections contribute to the diversity of the synthesized computational graphs.
However, simply putting them together will lead to invalid tests due to the undefined inputs for the operators in patterns.
To solve this, \tool{} designs two synthesis strategies, \ie, \textit{reusing the compatible inputs/outputs of existing nodes in contexts} and \textit{creating new nodes with compatible inputs/outputs}.
These strategies synthesize valid, optimization-aware computation graphs by leveraging connections between patterns and contexts.
The technical details are introduced in the following.

\subsubsection{Reusing Existing Node}
\label{sec:reuse_edge}
For each undefined input from the nodes in patterns, \tool{} prioritizes reusing compatible inputs/outputs of the existing nodes in the context to establish new connection edges, ensuring strong interactions between patterns and contexts.
Edge constraints in a computational graph are highly strict, requiring that the output of a node (\ie, a tensor variable) must align in type and shape with the input of the subsequent node.
In fact, it is less likely to find a compatible input/output of a node from the context that satisfies the strict constraints for the edge connecting an undefined input of the node in the pattern.
If no compatible input/output of nodes is found, 
\tool{} loosens the constraints by removing specific shape and type values 
and retaining only the intrinsic constraints of each node (\eg, the input of \mycode{Conv2D} must be a 4-dimensional tensor).
This abstraction increases synthesis flexibility without reducing optimization awareness, as DL compiler optimizations mainly depend on graph structure rather than the specific tensor values.
Figure~\ref{fig:motivation_example}(d) shows an abstract pattern derived from the pattern in Lines 5-8 of Figure~\ref{fig:motivation_example}(b). 
The abstract pattern eliminates the constraints for the input of the node \mycode{R.tensor\_to\_shape} (\ie, \mycode{R.Shape([2,4])}) in the concrete pattern, allowing a tensor with other shapes to be a valid input.

To merge the abstract pattern into the seed graph, \tool{} first randomly selects a synthesis point in the seed graph\change{.}{, which can improve the diversity of the generated tests.}
All outputs of nodes before the synthesis point form a preceding context, acting as a variable pool for repairing undefined inputs.
Specifically, for each undefined input in the abstract pattern, \tool{} searches the preceding context for compatible variables satisfying operator constraints
\add{(\eg, the inputs/outputs shape and type constraints)}. The operator constraints are reused from the source code of NNSmith~\cite{NNSmith}.
If a compatible variable is found, \tool{} establishes an edge connection between the seed graph and the abstract pattern by assigning the selected variable as the input of the node with an undefined input.
Meanwhile, to ensure that the context fully encompasses the abstract pattern, \tool{} further creates connections between the broken output edge in the pattern and the context.
Similarly, for each broken output edge, \tool{} searches from the succeeding context, a variable pool with all outputs of nodes after the synthesis point, for a compatible input of node in the seed graph and creates a connection by assigning the variable in the broken output edge as the input of the selected node with compatible input.
Finally, a valid optimization-aware computational graph is generated.

\begin{figure}
    \centering
    \includegraphics[width=\linewidth]{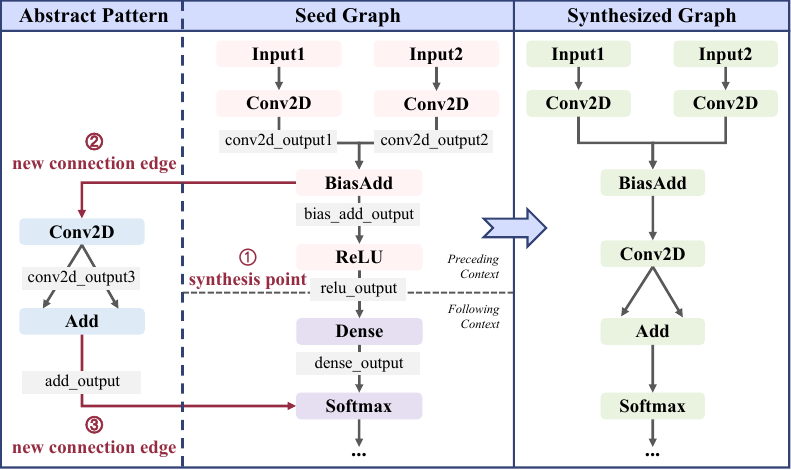}
    \caption{Fixing undefined inputs by reusing inputs/outputs of existing nodes in the context}
    \label{fig:synthesize}
\end{figure}

For example, in Figure~\ref{fig:synthesize}, 
\tool{} first randomly selects the synthesis point (\circled{1}) and classifies inputs/outputs of nodes in the seed graph into the preceding context or the succeeding context\del{ based on the synthesis point}.
Then, for the undefined input of nodes in the pattern (\ie, the input of the \mycode{Conv2D} operator in the abstract pattern), \tool{} selects a compatible variable (\eg, the \mycode{bias\_add\_output}\del{ that is a 4-dimension tensor}) from the preceding context and creates a new connection between them (\circled{2}).
Similarly, for the broken output edge \del{in the pattern} (\ie, \mycode{add\_output}), \tool{} \change{selects a compatible input}{links them to compatible inputs} (\del{\eg, the input }of the \mycode{Softmax}) in the succeeding context \del{and establishes a new connection between them} (\circled{3}).
Finally, the abstract pattern is well merged into the seed graph\del{ by reusing inputs/outputs of existing nodes}, generating a new synthesized computational graph.

\subsubsection{Creating New Nodes}
\label{sec:auxiliary_adding}
Not all undefined inputs in the pattern can find a compatible tensor variable in the seed graph, even in the abstract pattern with loosened constraints.
If no compatible tensor variable is found in the context, the aforementioned strategy of reusing existing nodes does not work.
In this situation, \tool{} creates new nodes in the seed graph as bridges
between the pattern and the seed graph.
Specifically, \tool{} randomly selects a tensor variable from the preceding context and compares its attributes (\ie, shape and type) with the constraints of the undefined input.
If the shapes are inconsistent, \tool{} first adjusts the element count (\ie, total number of the data point) of the selected tensor variable by inserting a \mycode{Padding} or \mycode{Cropping} node, generating a new tensor variable that satisfies the element count constraints for the undefined input.
Subsequently, \add{when the new tensor variable has different tensor dimension with the undefined input,} \tool{} appends a \mycode{Reshape} node to align the tensor dimensions (\ie, the number of axes) with those of the undefined input, thereby producing a shape-compatible tensor variable.
Additionally, when the types differ, \tool{} applies a \mycode{Retype} node to perform \add{explicit} type conversion \add{to the type of the undefined input}, ensuring the adjusted tensor variable satisfies the type inconsistency of the pattern.
Finally, the compatible tensor variable is assigned to the undefined input, forming a new connection edge and thus synthesizing a new computational graph. 
The same method is also used for fixing the broken output edge in the pattern.

\begin{figure}
    \centering
    \includegraphics[width=\linewidth]{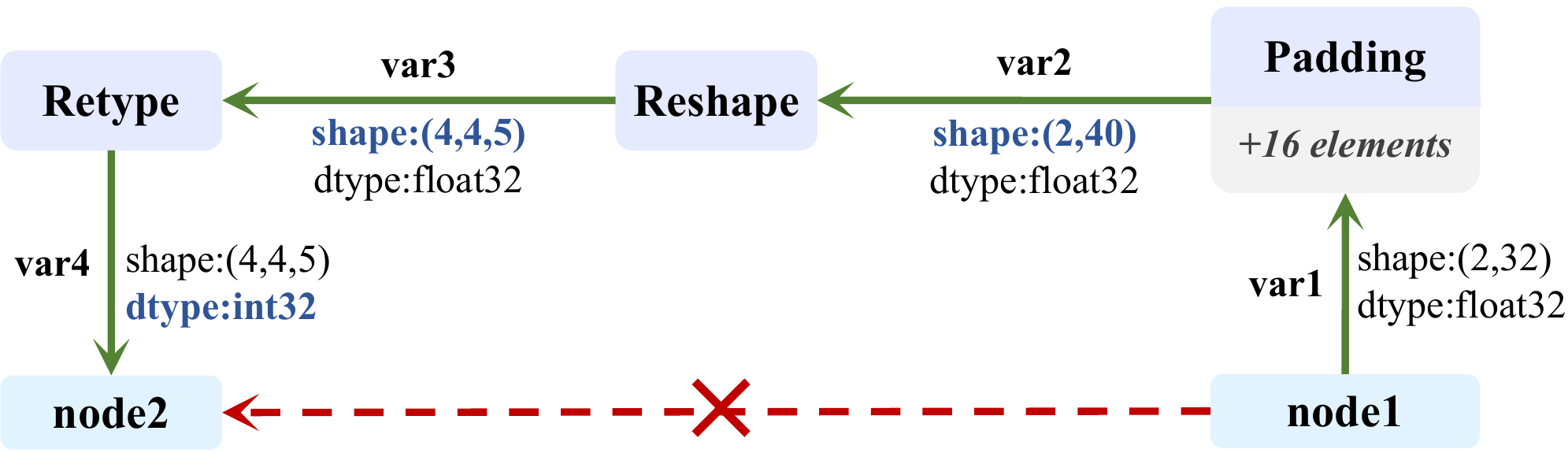}
    \caption{Fixing undefined inputs by creating new nodes}
    \label{fig:reshape_layer}
\end{figure}

Figure~\ref{fig:reshape_layer} shows an example of creating new nodes for fixing the undefined input \mycode{node2}\del{ with \tool{}}. 
\tool{} randomly selects a variable (\ie, \del{the output variable} \mycode{var1}\del{ of the \mycode{node1}}) from the context. 
The selected tensor variable \mycode{var1} has a shape of \mycode{(2,32)} and a type of \mycode{float32}.
However, the undefined input \mycode{node2} expects a tensor variable with shape \mycode{(4,4,5)} and type \mycode{int32} as input.
To align the shapes,
\del{ \tool{} first applies a \mycode{Padding} operator to add 16 elements for the \mycode{var1} and generate a new connection variable \mycode{var2} to meet the element size constraints.
Then, \tool{} appends a \mycode{Reshape} operator to adjust the dimension constraints of \mycode{var2} and output the shape-compatible variable \mycode{var3}.}
\add{\tool{} applies a \mycode{Padding} operator to create equal-sized \mycode{var2} followed by a \mycode{Reshape} operator to produce shape-compatible \mycode{var3}.
}
To further align the type constraint, \tool{} appends a \mycode{Retype} operator to convert the type of \mycode{var3} from \mycode{float32} into \mycode{int32} and obtain the type-compatible variable \mycode{var4}, which is also shape-compatible.
Finally, \tool{} establishes a new connection between the \mycode{node2} in pattern and the seed graph by assigning the \mycode{var4} as the input of \mycode{node2}.

\subsection{Test Oracle}
\label{sec:oracle}
A recent study found that most DL compiler bugs manifested as crashes and inconsistent inference results~\cite{shen2021comprehensive}.
Hence, we employ two test oracles in \tool{} to test DL compilers.

\textbf{Crash} refers to the unexpected termination of the compilation process. It is a widely-used test oracle in DL compiler testing~\cite{tzer,NNSmith,ma2022hirfuzz}.
According to a recent bug study in DL compilers~\cite{shen2021comprehensive}, bugs with crash symptoms occupy 59.37\% of all studied bugs, showing the significant importance of detecting crash bugs.
Note that crashes that produce error messages like ``unsupported'' are disregarded, as they usually indicate unsupported features rather than actual bugs.

\textbf{Inconsistency} refers to situations where a DL compiler yields inconsistent inference results for the same inputs with different optimizations or hardware devices (\eg, CPU and CUDA). 
Since floating-point operations in DL models always cause slight differences in results rather than a bug, we measure inference discrepancies using the widely-used Chebyshev distance~\cite{chebyshev} with a threshold of 1e-3 to \change{identify inconsistency bugs}{evaluate transformation equivalence in multi-layer networks, accounting for expected numerical error accumulation} following existing work~\cite{ma2022hirfuzz,NNSmith,comet}.
\add{This threshold is also consistent with industry practices, such as TVM’s official test suite for model optimization correctness validation.}
Since DL compilers are designed to maintain functional equivalence, any inconsistencies (exceeding the threshold) in inference results indicate a potential compiler bug.

If the synthesized computational graph passes testing without errors, \tool{} adds it to the seed pool for high-order synthesis. 
This enables combining patterns from different optimization-related tests, facilitating the creation of more complex optimization scenarios to uncover deeper optimization bugs.

\section{Evaluation}
\label{sec:evaluation}

\subsection{Setup}
\label{sec:setup}
Our study aims to address the following research questions (RQs).
\begin{itemize}
    \item \textbf{RQ1}: How does \tool{} perform in detecting previously unknown optimization bugs in DL compilers?

    \item \textbf{RQ2}: How does \tool{} compare against the state-of-the-art DL compiler fuzzers?
    \item \textbf{RQ3}: How does each key component in \tool{} contribute to the overall effectiveness?
    \item \textbf{RQ4}: How efficient does \tool{} in generating valid tests?
\end{itemize}


\subsubsection{Subjects}
\label{sec:subject}
Following recent DL compiler fuzzing works~\cite{NNSmith,ma2022hirfuzz,MT_DLComp}, we chose the widely-studied DL compilers, \ie, \textbf{TVM}~\cite{tvm} and \textbf{ONNXRuntime}~\cite{ort}, as the subjects in our study. 
TVM is an end-to-end compiler that can optimize and deploy DL models built on various DL frameworks across diverse platforms. 
It contains 24 block-level optimizations and 41 subgraph-level optimizations implemented across over 140 source files.
We chose the latest version of TVM (commit id: 292ecfd) as the subject under test to carry out the evaluation of \tool{}.
ONNXRuntime is a graph-optimized DL compiler designed for ONNX models, the most widely-adopted model format. Nearly all pre-trained DL models, developed using various frameworks (\eg, PyTorch~\cite{pytorch} and TensorFlow~\cite{tensorflow}), can be converted to ONNX format for executing diverse optimizations with ONNXRuntime. It includes over 21 block-level optimizations and 25 subgraph-level optimizations implemented across over 130 source files.
We chose the latest version of ONNXRunutime (commit id: 5c1b7cc) as the subject under test.
These DL compilers involve two representative computational graph formats, \ie, Relax IR and ONNX IR, which are compatible with nearly all mainstream DL compilers, including TVM~\cite{tvm}, MLC-LLM~\cite{mlc_llm}, ONNXRuntime~\cite{ort}, TensorRT~\cite{TensorRT}, OpenVINO~\cite{OpenVINO}, and Glow~\cite{glow}.
Hence, \tool{} implementation can be directly used for fuzzing other compatible DL compilers, indicating the high generalizability of our tool.

\subsubsection{Baselines}
\label{sec:baselines}
We compared \tool{} with two state-of-the-art DL compiler fuzzers (\ie, \textbf{NNSmith}~\cite{NNSmith} and \textbf{WhiteFox}~\cite{whitefox}) \add{as well as a custom LLM-based technique we developed (denoted as \textbf{\llmbaseline{}})}. 
NNSmith is a grammar-based computational graph generator that produces diverse and valid computational graphs for testing DL compilers. It ensures the validity of computational graphs through constraint solving. 
WhiteFox is a white-box fuzzer built upon LLMs. It utilizes an analysis LLM to extract optimization triggering conditions from source code and a generation LLM to create computational graphs aligned with these specific conditions.  
\add{\llmbaseline{} puts documented tests and corresponding optimization descriptions into the prompt and then instructs LLMs to generate tests to sufficiently test the optimization.
Building upon previous work~\cite{whitefox}, the method applies few-shot learning with identical configuration parameters to WhiteFox.
To evaluate the effectiveness of different LLMs in generating optimization-aware tests, we implemented this technique using two different state-of-the-art LLMs (\ie, DeepSeek-v3~\cite{deepseek} and Qwen2.5-Coder~\cite{qwen}), which are referred to as $\llmbaseline{}_{DS}$ and $\llmbaseline{}_{Qwen}$, respectively.
}
Other DL compiler testing techniques, either cannot test the compiler optimizations for computational graphs (\ie, OPERA~\cite{opera} and TZer~\cite{tzer}) or have been demonstrated less effective than NNSmith or WhiteFox  (\ie, MT-DLComp~\cite{MT_DLComp}, GenCoG~\cite{gencog}, and HirGen~\cite{ma2022hirfuzz}).
Therefore, we do not include them as baselines for comparison.

\subsubsection{Metrics}
We mainly targeted the following metrics for evaluation.
\label{sec:metric}
First, we used \textbf{the number of detected bugs} to measure the effectiveness of a DL compiler fuzzing technique following the existing work~\cite{ma2022hirfuzz,NNSmith,opera, suo2025desil,ma2025bounded}.
During the fuzzing process, a lot of test failures might be produced, but many of them might be caused by the same bugs.
Therefore, we performed de-duplication on them.
For crash bugs, we de-duplicated them based on the crash messages following the existing work~\cite{mlirsmith,opera,suo2024fuzzing}. 
For each inconsistency bug, we identified the minimal set of optimizations that could still reveal the bug based on the idea of delta debugging~\cite{zeller2002simplifying,wang2021probabilistic}.
If two inconsistency bugs had the same minimal sets of optimizations, we regarded them as duplicates.
Each distinct bug was submitted to developers for further diagnosis.
Here, we measured the number of submitted bugs after de-duplication, and the number of confirmed/fixed bugs.
Also, we used \textbf{code coverage} as another metric to assess \tool{}'s effectiveness following the existing work~\cite{NNSmith, mlirsmith,tzer,wu2023jitfuzz}.
Since \tool{} was designed for fuzzing DL compiler optimizations,  
we measured the line and branch coverage with regard to the source code of DL compiler optimizations.
To evaluate \tool{}'s efficiency in generating valid tests in RQ4, we measured \textbf{the number of valid tests} generated by it within a given fuzzing time budget.


\subsubsection{Implementations}
\label{sec:imple}

\tool{} collected optimization-aware patterns from the documented tests of TVM and ONNXRuntime (\ie, all tests in the folder of \texttt{test/tvm/relax/transform} for TVM and all tests in the folder of \texttt{onnxruntime/test/optimizer} for ONNXRuntime).
\add{Both TVM (65 optimizations) and ONNXRuntime (46 optimizations) achieve 100\% optimization coverage, with each optimization supported by multiple documented tests. 
}
In total, we obtained \TVMDonorNum{} and \ORTDonorNum{} patterns for TVM and ONNXRuntime, respectively.
\add{Since DL compilers (including TVM~\cite{tvm}, ONNXRuntime~\cite{ort}, OpenVINO~\cite{OpenVINO}, TorchInductor~\cite{pytorch2}, Glow~\cite{glow}, and nGraph~\cite{ngraph}) always provide well-documented tests for optimizations, the availability of such tests does not limit \tool{}.}

We then collected a set of tests generated by NNSmith as seed graphs for \tool{}.
Note that \tool{} is not specific to this source of seed graphs.
We selected NNSmith rather than WhiteFox as the seed provider since the latter required much more costs for generating tests due to heavily depending on LLMs.
For NNSmith, it is efficient to generate a large number of tests, and thus we collected 4k tests (taking only 10 minutes for generation) as seed graphs in our study, which could provide diverse contexts for test synthesis.
\add{All these seed graphs did not trigger bugs. 
}

We directly adopted the implementation of NNSmith released by the authors~\cite{NNSmith} for comparison.
For WhiteFox, \change{We}{we} adapted the code for \del{the} compatibility with TVM and ONNXRuntime. 
Furthermore, we adopted Gcov~\cite{gcov_ref} and Lcov~\cite{lcov_ref} to collect and analyze the line and branch coverage achieved by each technique.



\subsubsection{Process}
\label{sec:procedure}
In RQ1, we ran \tool{} for two weeks on each DL compiler to sufficiently evaluate its effectiveness in detecting previously unknown bugs.
In the remaining RQs, we ran each studied technique for 12 hours (including both test generation and execution time) on each compiler following the existing work~\cite{mlirsmith,opera} for fair comparison.
Particularly, we repeated each experiment for five times and aggregated the bug detection results for comparison, in order to reduce the influence of randomness involved in the fuzzing process.
\add{Note that OATest takes 4k tests generated by NNSmith (within 10 minutes) as seeds, while NNSmith ran for 12 hours, making comparison between them meaningful.}
All experiments were conducted on a server with an Intel Xeon Gold CPU, four A800 GPUs, and 500 GB of RAM, operating on a 64-bit Ubuntu 16.04 OS.


\subsection{RQ1: Previously Unknown Bugs Detected by \tool{}}
\label{sec:result_bugs}

\captionsetup[table]{labelfont=bf, labelsep=colon}
\begin{table}[]
    \centering
    \caption{Previously unknown Bugs found by \tool{}}
    \label{tab:bugs}
    \resizebox{\linewidth}{!}{
    \begin{tabular}{c|c|c|c|c|c}
\hline
\textbf{Compiler} & \textbf{Issue} & \textbf{Symptom} & \textbf{RC} & \textbf{Optimization} & \textbf{Status} \\ \hline
TVM & \#17120 & Crash & ICL & MergeCompositeFunctions & Fixed \\
TVM & \#17121 & Crash & TSP & DeadCodeElimination & Fixed \\
TVM & \#17175 & Crash & IEH & LegalizeOps & {\color[HTML]{000000} Fixed} \\
TVM & \#17176 & Crash & ICL & AttachGlobalSymbols & Fixed \\
TVM & \#17200 & Crash & ICL & LiftTransformParams*2 & Fixed \\
TVM & \#17205 & IC & ICL & RealizeVDevice*2 & Fixed \\
TVM & \#17210 & Crash & ICL & MergeCompositeFunctions & Fixed \\
TVM & \#17211 & Crash & TP & - & Fixed \\
TVM & \#17213 & Crash & ICL & RealizeVDevice*2 & Fixed \\
TVM & \#17215 & Crash & TSP & FlattenBuffer & Fixed \\
TVM & \#17217 & Crash & TSP & VMBuiltinLower & Fixed \\
TVM & \#17218 & Crash & TSP & {\color[HTML]{1F2328} VMBuiltinLower} & Fixed \\
TVM & \#17222 & Crash & ICL & DeadCodeElimination & Fixed \\
TVM & \#17223 & Crash & IEH & - & Confirmed \\
TVM & \#17231 & Crash & ICL & LiftTransformParams & Fixed \\
TVM & \#17235 & Crash & TP & - & Fixed \\
TVM & \#17242 & Crash & IEH & - & Fixed \\
TVM & \#17243 & Crash & TP & - & Fixed \\
TVM & \#17249 & IC & ICL & - & Fixed \\
TVM & \#17253 & IC & ICL & RemoveUnusedOutputs & Fixed \\
TVM & \#17254 & Crash & TP & ToMixedPrecision & Fixed \\
TVM & \#17269 & Crash & ICL & LazyTransformParams & Confirmed \\
TVM & \#17311 & Crash & IEH & - & Confirmed \\
TVM & \#17340 & Crash & ICL & \begin{tabular}[c]{@{}c@{}}KillAfterLastUse, \\ FoldConstant\end{tabular} & Confirmed \\
TVM & \#17341 & Crash & ICL & \begin{tabular}[c]{@{}c@{}}AnnotateTIROpPattern,\\ FuseOps, FuseTIR\end{tabular} & Confirmed \\
TVM & \#17348 & Crash & ICL & StaticPlanBlockMemory & Confirmed \\
TVM & \#17357 & Crash & ICL & \begin{tabular}[c]{@{}c@{}}FuseTIR, LambdaLift, \\ AllocateWorkspace\end{tabular} & Confirmed \\
TVM & \#17370 & Crash & IEH & LegalizeOps & Fixed \\
TVM & \#17483 & Crash & ICL & LegalizeOps & Fixed \\
TVM & \#17486 & Crash & ICL & LegalizeOps & Fixed \\
TVM & \#17487 & Crash & ICL & - & Confirmed \\
TVM & \#17488 & Crash & ICL & StaticPlanBlockMemory & Fixed \\
ORT & \#23086 & Crash & IEH & FuseQuickGeLU & Confirmed \\
ORT & \#23114 & Crash & ICL & FusedConv & Confirmed \\
ORT & \#23116 & Crash & TP & FuseReluClip & Confirmed \\
ORT & \#23118 & Crash & ICL & DeadCodeElimination & Confirmed \\
ORT & \#23119 & Crash & IEH & FusedConv & Confirmed \\
ORT & \#23138 & Crash & ICL & GeluFusion & Confirmed \\
ORT & \#23143 & IC & ICL & MemCpy & Confirmed \\
ORT & \#23199 & IC & ICL & Quantizer, Dequantizer & Confirmed \\
ORT & \#23258 & Crash & ICL & Dequantizer & Fixed \\
ORT & \#23284 & IC & ICL & - & Confirmed \\ \hline
\end{tabular}
}
\footnotesize
\vspace{1mm}
\\
\textbf{RC}: Root Cause; 
\textbf{ORT}: ONNXRuntime; 
\textbf{IC}: Inconsistency;
\textbf{ICL}: Incorrect Code Logic; 
\vspace{-1mm}
\textbf{TSP}: Tensor Shape Problem; 
\textbf{IEH}: Incorrect Exception Handling;
\textbf{TP}: Type Problem;

\end{table}

During two-week fuzzing, \tool{} detected \TotalBugsNum{} previously unknown bugs, of which \ConfirmedBugsNum{}/\FixedBugsNum{} have been confirmed/fixed by developers. 
The remaining bugs are still awaiting developers' feedback.
Among the \TotalBugsNum{} bugs, \TotalTVMBugsNum{} are TVM bugs and \TotalORTBugsNum{} are ONNXRuntime bugs. 
Table~\ref{tab:bugs} shows the details of all confirmed bugs detected by \tool{}, where Columns 1-6 represent the buggy compiler, the issue ID, the bug symptom, the root cause of the bug, the buggy optimization, and the bug status labeled by developers, respectively.
Particularly, some of our reported bugs are regarded as ``high-quality'' by the TVM community and our contributions have been appreciated~\cite{issue_17231,issue_17211, issue_17235}. 
These confirmed bugs cover a wide range of buggy optimizations, symptoms, and root causes. 
In the following, we analyzed them from the three aspects.

\begin{figure}
    \centering
    \includegraphics[width=\linewidth]{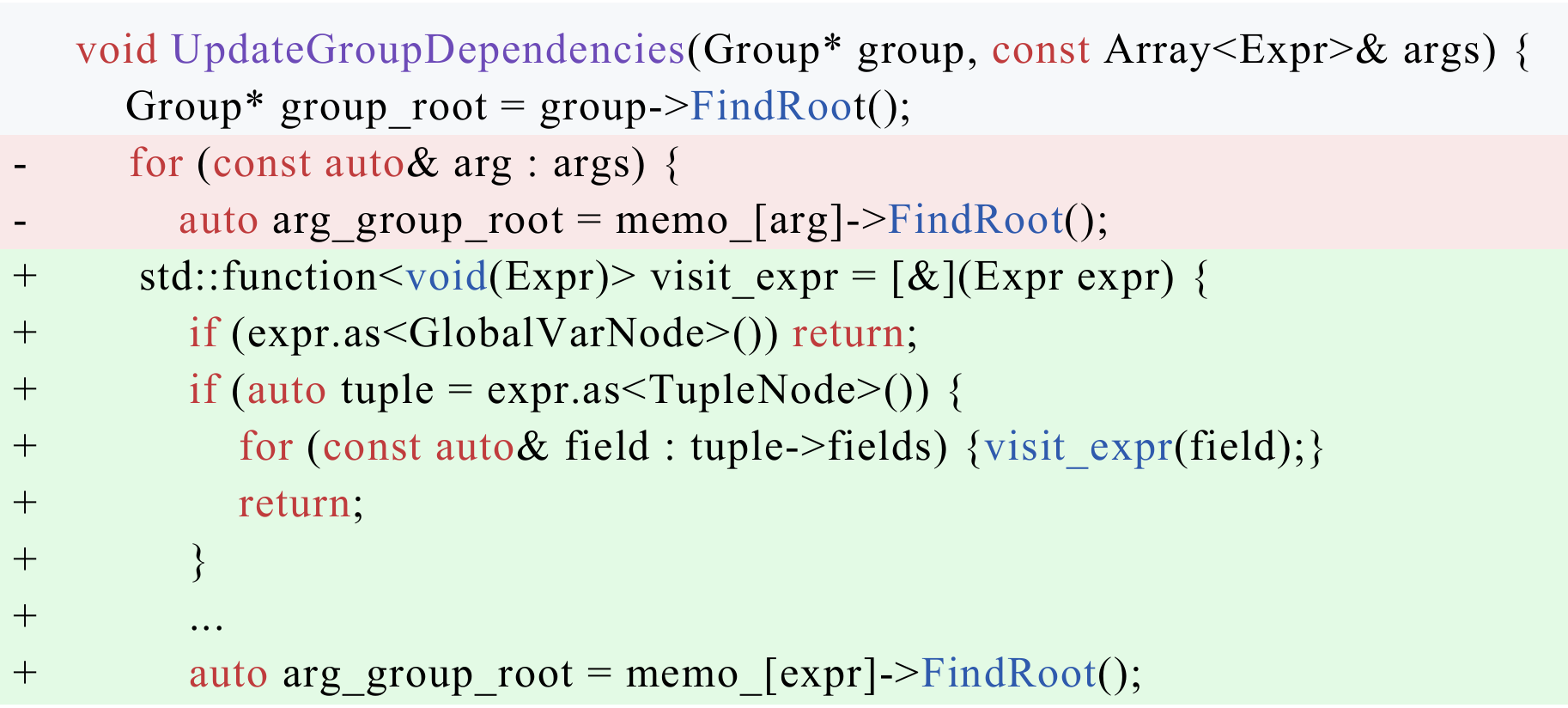}
    \caption{A crash bug detected by \tool{} (\#17120). The bug is caused by Incorrect Code Logic}
    \label{fig:segfault}
\end{figure}

\smallskip
\textbf{Buggy Optimization Analysis:}
As \tool{} is designed for fuzzing DL compiler optimizations, we first investigated whether each confirmed bug is an optimization bug and labeled the corresponding buggy optimizations based on the feedback from developers. 
From this table, \textbf{\OPTBugsNum{} out of \ConfirmedBugsNum{} confirmed bugs detected by \tool{} are optimization bugs}, including \TVMOPTBugsNum{} TVM bugs and \ORTOPTBugsNum{} ONNXRuntime bugs.
The \OPTBugsNum{} optimization bugs involve 26 optimizations, including 28 bugs involving a single optimization and 5 bugs involving multiple optimizations.
The remaining 9 bugs stem from flawed IR design in the optimization stage, not faulty optimizations. 
Since IRs serve as the object for optimization, identifying such flaws is also crucial.
The results demonstrate the effectiveness of \tool{} in uncovering previously unknown optimization bugs.



For example, a test created by \tool{} exposes an inconsistency bug~\cite{issue_pass} specifically when the \mycode{RealizeVDevice} \textit{pass} is executed twice.
Due to extensive code changes, this bug is not shown as a figure in this paper and more details can be found in the issue~\cite{issue_pass}.
Specifically, the \mycode{RealizeVDevice} \textit{pass} illegally mutated the relax expression in place. Consequently, applying \mycode{RealizeVDevice} a second time caused unintended updates in unrelated expressions, resulting in incorrect IRs.
To address this, a complex patch\cite{issue_pass} was introduced, refactoring the \mycode{RealizeVDevice} \textit{pass} to eliminate all in-place mutations 
with more than 600 lines of code changes.

\smallskip
\textbf{Symptom Analysis:}
The symptoms of our detected bugs include crashes
and inconsistency.
\textit{Crashes} indicate unexpected termination during the compilation process of DL compilers, with \TotalConfirmedCrashBugs{} out of the \ConfirmedBugsNum{} bugs exhibiting this symptom. 
\textit{Inconsistency}, observed in \TotalConfirmedWrongBugs{} confirmed bugs, refers to producing inconsistent inference results or IRs during the differential testing process. 
These bugs pose significant risks as they can \textit{silently} produce incorrect optimization results.
The results demonstrate the effectiveness of both test oracles in \tool{}.

Figure~\ref{fig:segfault} shows a crash bug in TVM~\cite{issue_segfault}.
A computational graph synthesized by \tool{} triggered this bug since it contains the optimization-aware pattern (\ie., fused operators) for activating the \mycode{MergeCompositeFunctions} \textit{pass} and another bug-triggering condition (\ie, \mycode{R.call\_tir} operator) provided by context.
The lack of checks for the \mycode{relax::GlobalVar} referenced by \mycode{R.call\_tir} in the \mycode{MergeCompositeFunctions} \textit{pass} leads to null pointer access, resulting in a crash.
This bug was fixed by adding type checks to avoid null pointer dereference~\cite{pr_segfault}.
However, detecting this bug is challenging for NNSmith and WhiteFox. 
NNSmith rarely triggers the \mycode{MergeCompositeFunctions} \textit{pass} through random grammar-based fuzzing, while WhiteFox struggles to generate a valid computational graph for triggering this \textit{pass}.


\smallskip
\textbf{Root Cause Analysis:}
According to the discussion from developers and the patch for the fixed bugs, we identified four distinct root causes behind the \ConfirmedBugsNum{} confirmed bugs detected by \tool{}, classified following a recent study on DL compiler bugs~\cite{shen2021comprehensive}.
The results demonstrate that \tool{} is capable of revealing bugs with diverse root causes, including Incorrect Code Logic, Tensor Shape Problem, Incorrect Exception Handling, and Type Problem.

\textit{Incorrect Code Logic} refers to the bugs arising from the incorrect algorithm logic (\eg, operator fusion),
requiring modification with many statements. 
26 bugs (61.90\% of the confirmed/fixed bugs) fall into this category.
Depending on the bug location, this category is further divided into 
\textit{Incorrect Optimization Code Logic} and \textit{Incorrect Non-Optimization Code Logic}. 
The former represents a specialized type of optimization bugs caused by incorrect optimization code logic, with 20 bugs belonging to this subcategory, accounting for 60.61\% of optimization bugs.
The results demonstrate that most of the bugs detected by \tool{} have broad harm to the DL compiler, especially for the optimization code.

\textit{Tensor Shape Problem} involves the bugs caused by wrong tensor shape operation, such as tensor shape inference and tensor shape transformation. 
Tensor shape
plays a vital role in DL compiler optimizations. 
4 confirmed bugs belong to this category.
\begin{figure}
    \centering
    \includegraphics[width=.95\linewidth]{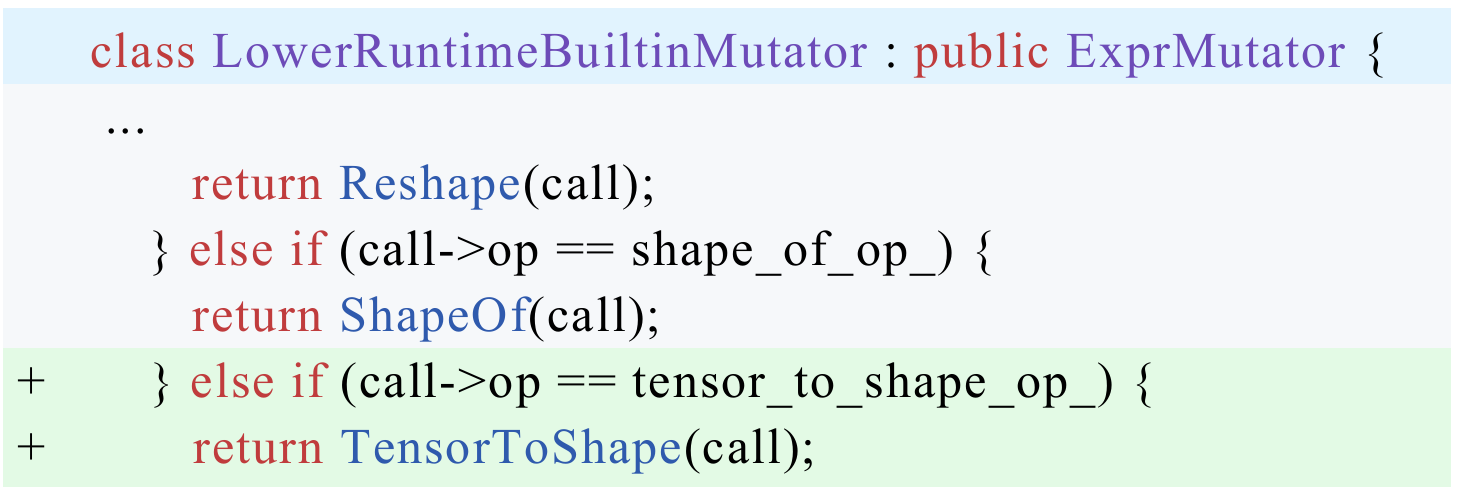}
    \caption{
        A crash bug detected by \tool{} (\#17217). The bug is caused by Tensor Shape Problem
    }
    \label{fig:bug_shape}
\end{figure}
Figure~\ref{fig:bug_shape} shows a Tensor Shape Problem bug~\cite{issue_shape}.
The \mycode{ShapeStructInfo} type is an acceptable input type for the \mycode{shape} argument of the \mycode{Reshape} operator.
However, the \mycode{Reshape} operator mistakenly required the \mycode{shape} argument to be either an inline \mycode{relax::ShapeExpr} or a variable bound to one. 
A test generated by \tool{} including a \mycode{Reshape} operator with the \mycode{shape} argument generated by \mycode{tensor\_to\_shape\_op}, triggered this bug and crashed unexpectedly. 
After reporting to developers, it was fixed with a patch~\cite{pr_shape} that applied the \mycode{TensorToShape} operation to support reshaping for the \mycode{ShapeStructInfo} type.

\textit{Incorrect Exception Handling} refers to the wrong management on exception. 7 confirmed bugs belong to this category.
Bugs in this category pose risks for compilers as they facilitate side channel attacks~\cite{d2015correctness}, enabling attackers to deduce the internal state of a computation without direct access.

\textit{Type Problem} refers to bugs involving type-related problems, such as type conversion and type inference. This category includes 5 confirmed bugs: 3 stemming from the tensor type problem, and 2 caused by the conventional data type problem.

\subsection{RQ2: Comparison with State-of-the-art Techniques}
\label{sec:result_coverage}

We then compared \tool{} with two state-of-the-art DL compiler fuzzing techniques (\ie, NNSmith and WhiteFox) and a custom LLM-based technique we developed (\ie, \llmbaseline{}).
Under the same setup (presented in Section~\ref{sec:procedure}), \tool{} detected \TotalTVMBugsNumCC{} (including 22 optimization bugs) and \TotalORTBugsNumCC{} (including 8 optimization bugs) previously unknown bugs in TVM and ONNXRuntime respectively, whereas NNSmith detected \NNSmithBugsNum{} (including 2 optimization bugs) and 0 bugs \change{and}{,} WhiteFox detected \WhiteFoxBugsNum{} (including 3 optimization bugs) and 1 (including 0 optimization bugs) bugs,
\add{
$\llmbaseline{}_{\textit{DS}}$ detected 4 (including 2 optimization bugs) and 2 (including 1 optimization bugs) bugs, and $\llmbaseline{}_{\textit{Qwen}}$ detected 5 (including 2 optimization bugs) and 2 (including 1 optimization bugs) bugs,}
respectively.
Among those bugs detected by baselines, only three are not detected by \tool{}, all of which occur in the model loading stage rather than compiler optimizations. 
The results demonstrate the superiority of \tool{} over \change{both}{four} baselines in DL compiler fuzzing, particularly in detecting optimization bugs.

\begin{figure}[t] 
    \subfigure{
        \includegraphics[width=0.475\linewidth]{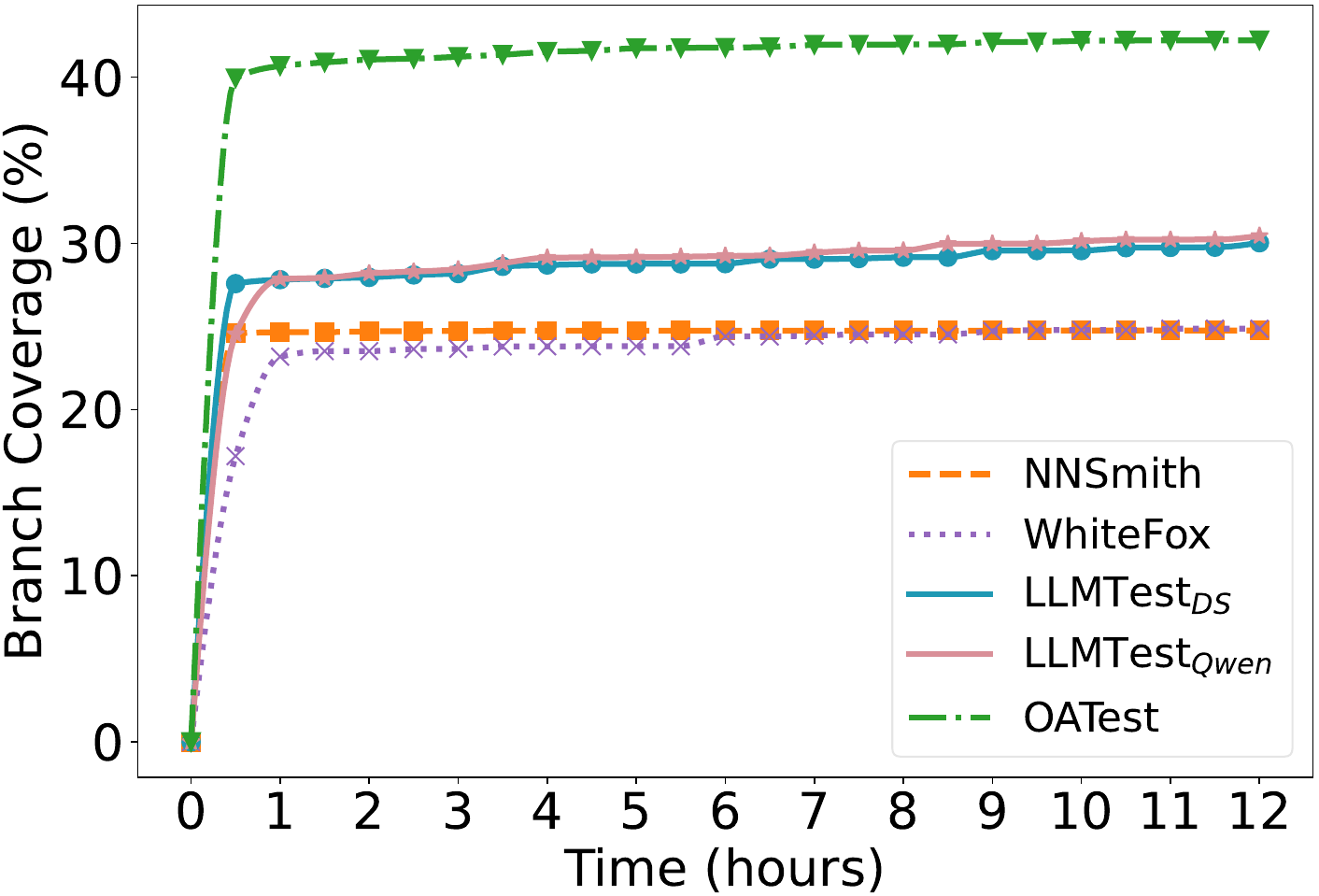}
        \label{fig:trend_branch}
    }
    \hfill
    \subfigure{
        \includegraphics[width=0.475\linewidth]{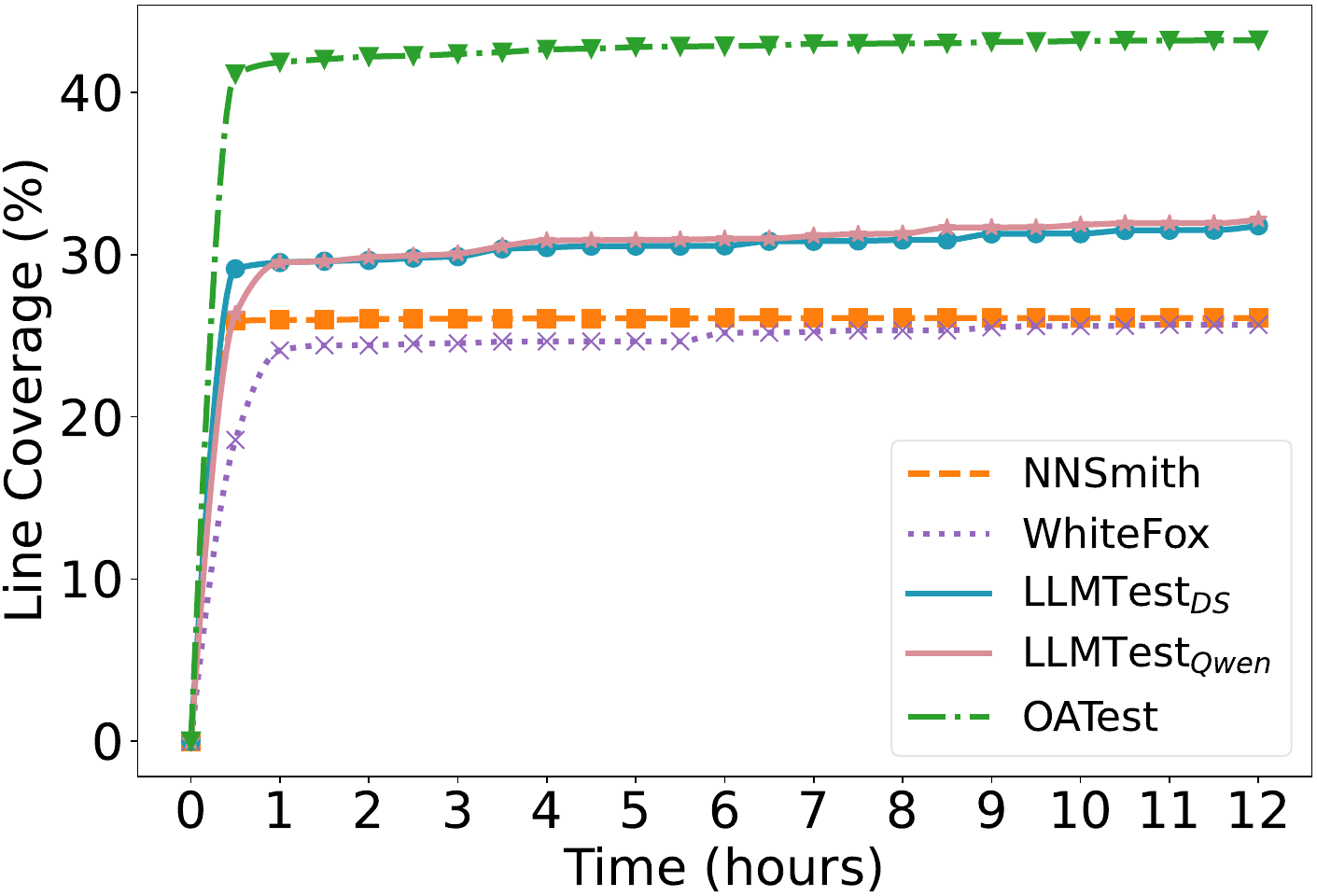} 
        \label{fig:trend_line}
    }
    \vspace{-3mm}
    \caption{\add{Branch and line coverage of TVM}}
    \vspace{-2mm}
    \label{fig:cov_tvm_trend} 
\end{figure}

\begin{figure}[t]
    \subfigure{
        \includegraphics[width=0.475\linewidth]{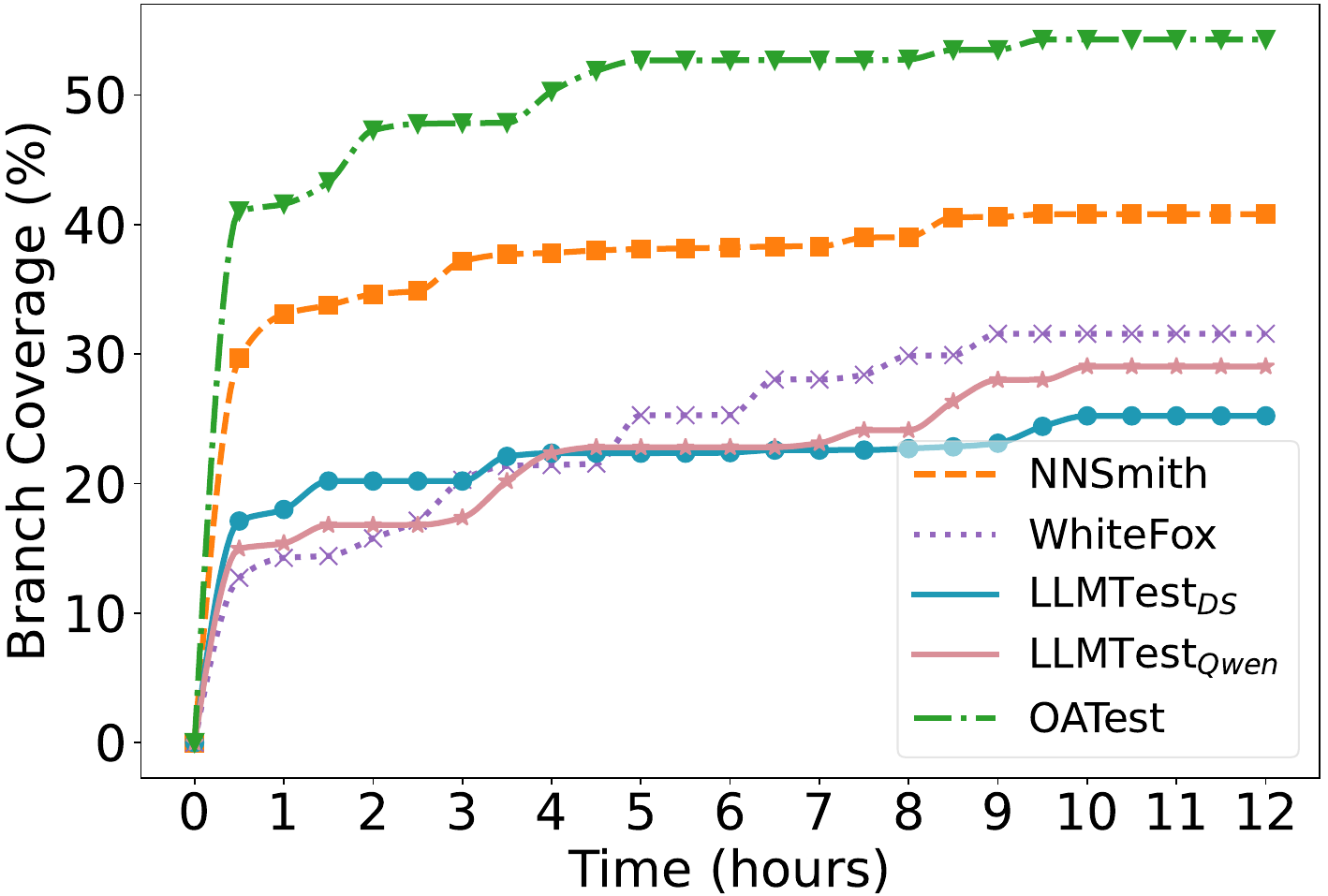}
        \label{fig:trend_ort_branch}
    }
    \hfill
    \subfigure{
        \includegraphics[width=0.475\linewidth]{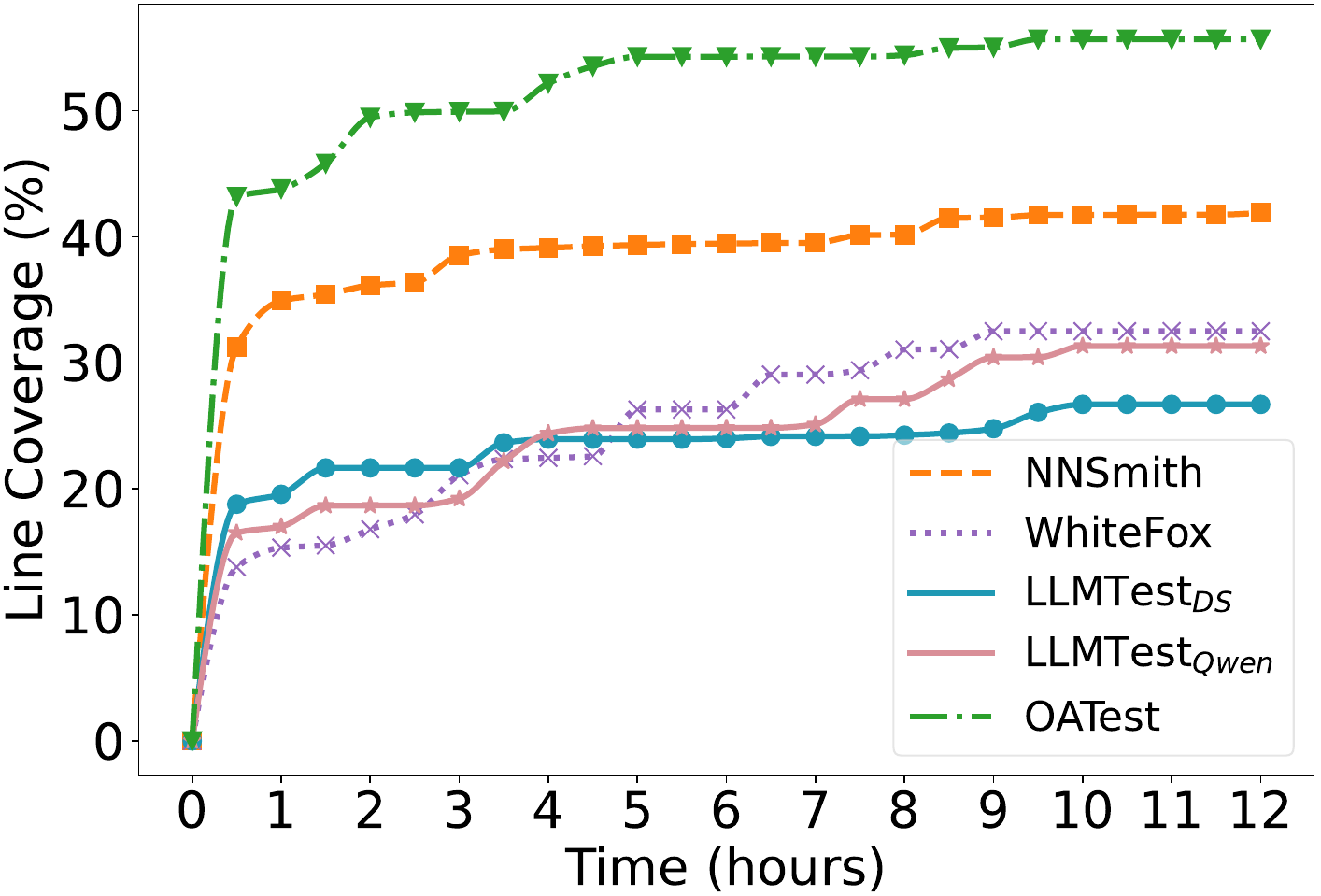} 
        \label{fig:trend_ort_line}
    }
    \vspace{-3mm}
    \caption{\add{Branch and line coverage of ONNXRuntime}}
    \vspace{-3mm}
    \label{fig:cov_ort_trend} 
\end{figure}

To further understand the superiority of \tool{} in bug detection, we compared the line coverage and branch coverage achieved by each studied technique.
Figures~\ref{fig:cov_tvm_trend} and~\ref{fig:cov_ort_trend} present the coverage growth within 12 hours on TVM and ONNXRuntime, respectively.
From these figures, \tool{} consistently achieves higher branch and line coverage than \change{both}{four} baselines on both TVM and ONNXRuntime. 
For example, compared to the more effective baseline in terms of code coverage (\ie, \change{NNSmith}{$\llmbaseline{}_{\textit{Qwen}}$ in TVM and NNSmith in ONNXRuntime}), \tool{} covers \change{70.55\%}{32.21\%} and 32.29\% more branches as well as 
\change{65.66\%}{32.29\%} and 32.89\% more lines in the optimization code for TVM and ONNXRuntime, respectively.
That is, it outperforms the more effective baseline by \change{60.20\%}{\BranchcovImporve{}} in branch coverage and \change{66.98\%}{\LinecovImporve{}} in line coverage on average. 
The results demonstrate the capability of \tool{} in improving code coverage, thereby contributing to its significant bug detection effectiveness.

\add{
While WhiteFox, {$\llmbaseline{}_{\textit{DS}}$}, and {$\llmbaseline{}_{\textit{Qwen}}$} are designed to test optimization in DL compilers, they exhibit limited bug detection capabilities and poor code coverage. Different LLM implementations of \llmbaseline{} achieve similar effectiveness, and employing alternative LLMs does not significantly improve performance. Our analysis traces these shortcomings to two key factors: low validity rates and inefficiency, which we further examine in Section~\ref{sec:test_quality}.
}

\subsection{RQ3: Contribution of Each Key Component}
\label{sec:result_ablation}

\add{
\subsubsection{Contribution of Different Seed Graphs}
\label{sec:ablation_seeds}

\begin{figure}[h!]
    \centering
    \includegraphics[width=0.85\linewidth]{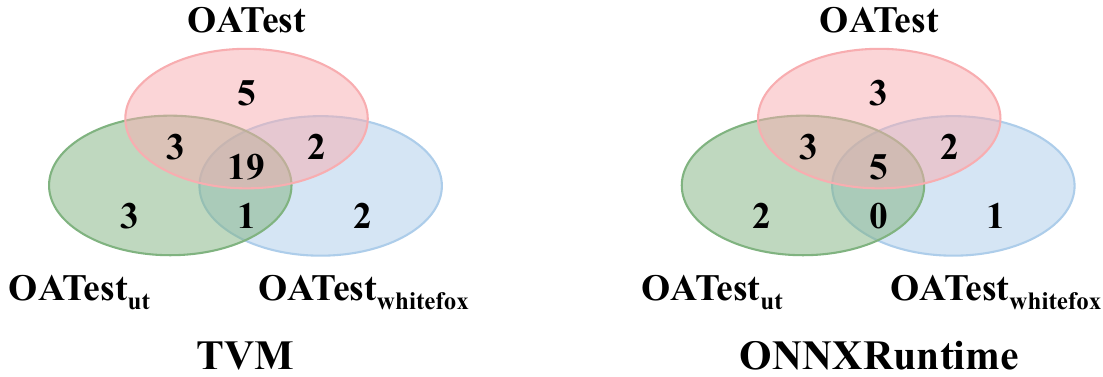}
    \caption{Venn graph of bugs detected by different techniques}
    \label{fig:seed_vn}
\end{figure}

To assess the influence of seed graphs, we designed two variants of \tool{}: OATest$_{dt}$, which uses \textbf{d}ocumented \textbf{t}ests as seed graphs, and OATest$_{whitefox}$, which relies on tests generated by WhiteFox as seeds.
Both variants used the same number of seed graphs (\ie, 4k) as \tool{} (using NNSmith-generated tests as seeds).
Figure~\ref{fig:seed_vn} shows a Venn diagram comparing the bugs detected by the three methods with different seeds under the same experimental configuration as RQ2. 
\tool{} detected 29 bugs in TVM and 13 in ONNXRuntime. 
In comparison, OATest$_{\textit{dt}}$ found 26 and 10 bugs, while OATest$_{\textit{whitefox}}$ identified 24 and 8 bugs, respectively.
Note that each technique detected several unique bugs missed by the others. 
Also, all of them outperform three baselines, \ie, NNSmith (detected 5 and 0 bugs), WhiteFox (detected 3 and 1 bugs), and two implementations of \llmbaseline{} (detected 4 and 2 bugs in DeepSeek implementation and detected 5 and 2 bugs in Qwen implementation). 
These results demonstrate the effectiveness of our approach in bug detection, regardless of the seeds used, and suggest that diversifying seed graphs could further improve bug-finding capabilities.

}

\subsubsection{Contribution of Pattern Extraction}
\label{sec:ingredient_type}

\begin{table}[]
\centering
\caption{
Bugs detected by OATest and variants}
\label{tab:variant_bugs}
\resizebox{\linewidth}{!}{
    \begin{tabular}{@{}cccccc@{}}
        \toprule
         & $\tool{}_{\textit{block}}$ & $\tool{}_{\textit{subgraph}}$ & $\tool{}_{\textit{graph}}$ &
         $\tool{}_{\textit{noopt}}$ &
         \tool{} \\ \midrule
        TVM & 14 & 20 & 11 & 7 & \TotalTVMBugsNumCC{} \\
        ONNXRuntime & 8 & 7 & 5 & 3 & \TotalORTBugsNumCC{} \\ \midrule
        \textbf{Total} & 22 & 27 & 16 & 10 & \TotalBugsNumCC{} \\ \bottomrule
    \end{tabular}
}
\end{table}

\add{
To evaluate our pattern extraction strategy (\ie, adaptively extracting block-level or subgraph-level patterns from a \pair{CG, \textit{pass}} pair based on the characteristics of \textit{pass}), we first tracked how often optimization passes linked to injected patterns were triggered during test execution. 
Specifically, for each test generated by \tool{}, we executed it and collected the coverage on the source code of the target optimization.
During the 12-hour fuzzing period (see Section~\ref{sec:procedure}), 91.36\% and 75.49\% of the tests synthesized by \tool{} successfully triggered the target optimization in TVM and ONNXRuntime, respectively.
A small portion of tests generated by \tool{} fail to trigger the target optimization due to the random pattern injection.
In comparison, WhiteFox, another optimization-aware test generation technique, achieved substantially lower optimization-triggering rates (\ie, 27.79\% in TVM and 21.52\% in ONNXRuntime) than \tool{}, demonstrating \tool{}'s effectiveness in pattern extraction.
}

\change{We first investigated the contribution of our pattern extraction strategy (\ie, adaptively extracting block-level or subgraph-level patterns from a \pair{CG, \textit{pass}} pair according to the characteristics of \textit{pass})}{Additionally, we further investigated the contribution of our pattern extraction strategy} by constructing four variants of \tool{}:
(1) \textbf{$\tool{}_{\textit{block}}$} always extracts block-level patterns from each \pair{CG, \textit{pass}} pair, 
(2) \textbf{$\tool{}_{\textit{subgraph}}$} always extracts subgraph-level patterns from each \pair{CG, \textit{pass}} pair, and 
(3) \textbf{$\tool{}_{\textit{graph}}$} does not extract fine-grained patterns from a \pair{CG, \textit{pass}} pair but directly use the entire graph (\ie, \textit{CG}) for test synthesis.
(4) \textbf{$\tool{}_{\textit{noopt}}$} randomly extracts the same number of patterns as \tool{} from the optimization-unrelated tests documented in DL compilers.
\add{Each variant adopts a fixed pattern extraction strategy instead of the granularity-aware strategy used in \tool{}.}

Table~\ref{tab:variant_bugs} presents the number of bugs detected by \tool{} and each variant during the same fuzzing time budget (introduced in Section~\ref{sec:procedure}).
By comparing \tool{} with $\tool{}_{\textit{block}}$ and $\tool{}_{\textit{subgraph}}$, our technique detected more bugs (\TotalBugsNumCC{}) than the two variants (22 and 27), demonstrating the contribution of our adaptive pattern extraction strategy.
By comparing $\tool{}_{\textit{graph}}$ with $\tool{}_{\textit{block}}$, $\tool{}_{\textit{subgraph}}$ and \tool{}, the former detected the smallest number of bugs in both TVM and ONNXRuntime.
Specifically, $\tool{}_{\textit{graph}}$ detected only 11 TVM bugs and 5 ONNXRuntime bugs.
The results demonstrate the contribution of extracting fine-grained patterns for test synthesis, compared to using the entire computational graph from a documented test.
By comparing $\tool{}_{\textit{noopt}}$ with \tool{}, the latter detects much more bugs (\TotalBugsNumCC{}) than the former (10) in total.
Furthermore, \tool{} covered 51.24\% more branches and 47.33\% more lines in the optimization code than $\tool{}_{\textit{noopt}}$ on average.
These results demonstrate that the extracted patterns by \tool{} are more optimization-aware, allowing it to detect more optimization bugs.

\begin{figure}
    \centering
    \includegraphics[width=.9\linewidth]{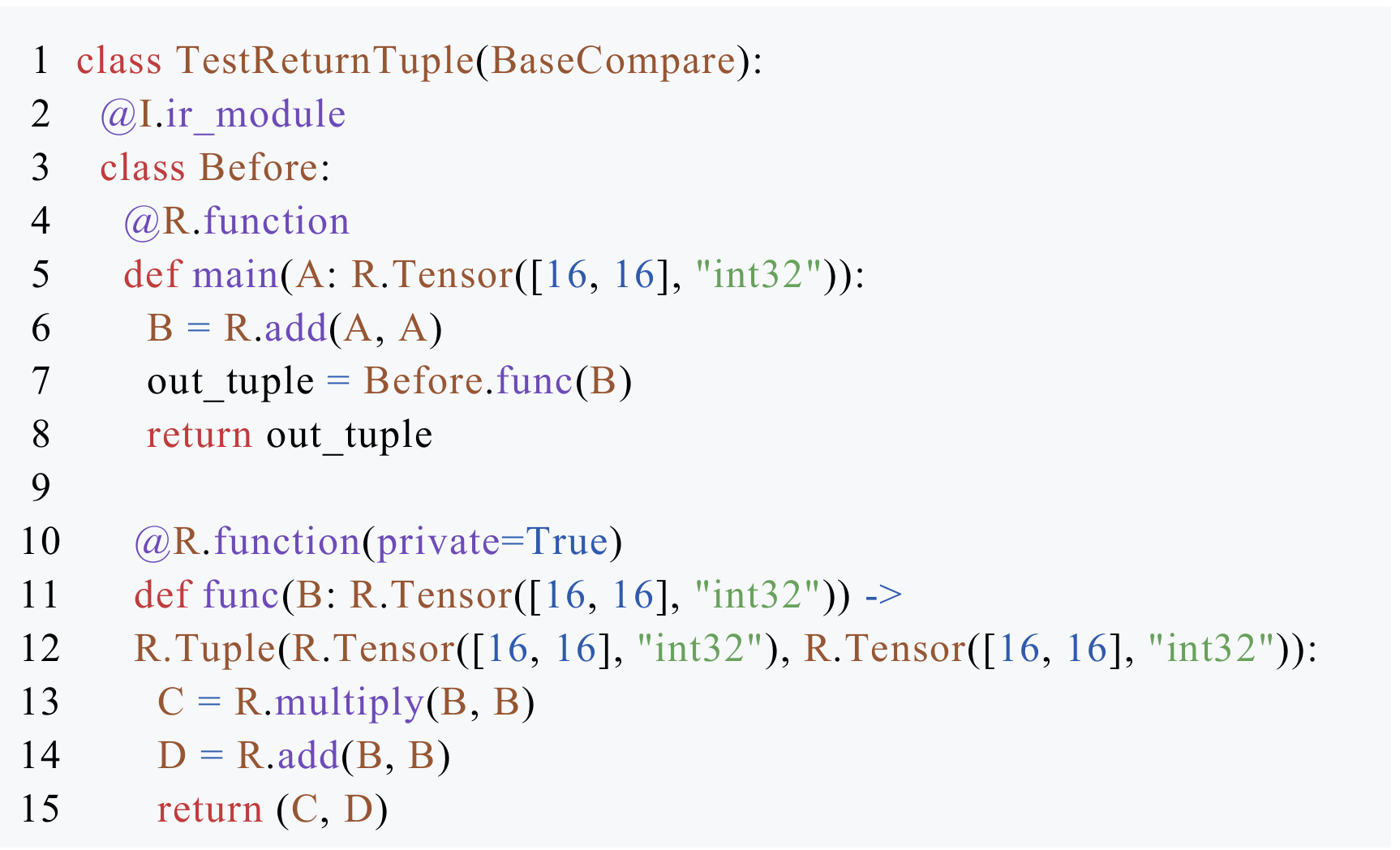}
    \caption{A bug-revealing computational graph synthesized with subgraph-level pattern}
    \vspace{-2mm}
    \label{fig:bug_subgraph}
\end{figure}

Figure~\ref{fig:bug_subgraph} shows a reduced bug-revealing test synthesized by \tool{} at the subgraph level.
The \mycode{main} function is a seed graph and the \mycode{func} function is a subgraph-level pattern.
Applying this \textit{pass} to the synthesized computational graph produced an incorrect IR silently.
\change{According to developers' feedback,  t}{T}he \mycode{RemoveUnusedOutputs} \textit{pass} mistakenly identified a tuple element as \change{being used exclusively}{only} in \mycode{TupleGetItem} nodes, failing to account for scenarios where tuples are returned from a function.
The bug can only be triggered when a function calls a subroutine that creates a tuple and then immediately returns that tuple, making it detectable only through subgraph-level synthesis. 
Block-level synthesis tests fail to trigger the bug \change{due to missing subgraph context}{because they lack the necessary subgraph structure} (\ie, the \mycode{func} function).
Furthermore, even when the entire seed graph is used as a synthesis pattern, the bug remains undetected because the tuple produced by the subgraph is not propagated as the output of the overall graph.
To fix it, the developers committed a patch~\cite{pr_subgraph} by \change{correcting the implementation of}{modifying} \mycode{RemoveUnusedOutputs} \textit{pass} to check for tuple object usage, 
not just for the \mycode{TupleGetItem}.

\subsubsection{Contribution of Test Synthesis}



We then investigated the contribution of test synthesis (\ie, synthesizing patterns into seed graphs).
First, all the documented tests for optimization (used for pattern extraction) and seed graphs (served as context) used by \tool{} cannot detect any bugs in both TVM and ONNXRuntime, but synthesizing them with \tool{} detected a number of bugs presented before, demonstrating the contribution of test synthesis in bug detection.
Then, we collected the line coverage and branch coverage achieved by all the documented tests and seed graphs, and analyzed whether \tool{} can achieve more optimization code than them.
Within the given fuzzing time budget, 
\tool{} further covered 3.75\% new branches and 3.57\% new lines in the optimization code of TVM, and covered 5.95\% new branches and 5.22\% new lines in the optimization code of ONNXRuntime, compared to all the used documented tests and seed graphs, respectively.
The results further confirm the contribution of test synthesis with \tool{}.

\subsection{RQ4: Efficiency}
\label{sec:test_quality}
Within the given fuzzing time budget, \tool{} generated and executed 41.9k and 112k valid tests for TVM and ONNXRuntime, respectively, while NNSmith generated and executed 37.2k and 103.7k valid tests for TVM and ONNXRuntime, \del{and }WhiteFox generated and executed 1.6k and 2k valid tests\add{, $\llmbaseline{}_{\textit{DS}}$ generated and executed 1.7k and 1.9k valid tests, and $\llmbaseline{}_{\textit{Qwen}}$ generated and executed 2.6k and 3.1k valid tests}. 
The results demonstrate the efficiency superiority of test synthesis with \tool{} over grammar-based test generation with NNSmith and LLM-based test generation with WhiteFox and \llmbaseline{}.
Through further analysis, the test generation process consumed less than 10\% of the entire fuzzing time for \tool{}, but over 95\% for WhiteFox\add{ and \llmbaseline{}}.
This indicates that the efficiency of test generation is a major bottleneck for WhiteFox \add{and \llmbaseline{} }due to LLMs' hallucinations, causing low test validity (37\% for TVM, 44.36\% for ONNXRuntime in WhiteFox \add{and 42.71\% for TVM, 44.54\% for ONNXRuntime average in two different implementations of \llmbaseline{}}) and limiting its efficiency.

\section{Discussion}
\label{sec:discuss}

\subsection{Threats to Validity}
\label{sec:threats}
The threat to internal validity mainly lies in the implementation of \tool{} and the adaptation of WhiteFox for TVM and ONNXRuntime. 
To mitigate this, two authors thoroughly reviewed and tested the code.
For WhiteFox, we adapted its code for compatibility with TVM and ONNXRuntime, following 
its original paper. 
To reduce GPT-4~\cite{gpt4} costs in WhiteFox, we substituted it with the open-source Phind-CodeLlama-34B-v2~\cite{phind} as the \textit{analysis LLM} and verified its correctness by reproducing its original evaluation results.

The main threat to external validity arises from the choice of DL compilers. We selected TVM and ONNXRuntime, two widely adopted DL compilers.
The two DL compilers are particularly representative due to their computational graph formats: Relax IR and ONNX IR, which are compatible with nearly all mainstream DL compilers (as illustrated in Section~\ref{sec:subject}).
Their broad compatibility ensures that the tests generated by \tool{} can be effectively adopted for testing other DL compilers, underscoring \tool{}'s high generalizability and reducing this threat.


The threat to construct validity mainly arises from the randomness. To reduce this threat, we repeated all experiments involving randomness for five times and reported their aggregated results. 

\add{
\subsection{Optimization-based Fusion Operators}
Optimization-based Fusion Operators (OBFOs) are crucial in DL compilers, boosting computational efficiency by reducing redundancy and kernel launch overhead.
Since optimizations in DL compilers are context-aware, applying fusion patterns across different contexts can expose new optimization opportunities. 
Currently, most optimization patterns are handcrafted by experts, yielding complex fusion algorithms prone to bugs.
For example, 5 of 10 confirmed ONNXRuntime bugs detected by \tool{} occurred in OBFOs.
Looking ahead, research in OBFOs should pursue more generalized, automated, and robust approaches.
Promising avenues include:
(1) Unified ML-driven fusion, which employs learned cost models to generate adaptive fusion strategies, simplifying implementation 
logic compared to maintaining separate complex algorithms for each optimization, thus reducing bug-prone code;
(2) Context-aware fusion mechanisms, which iteratively integrate existing optimization patterns with new operators in the context (like high-order synthesis strategy in \tool{}) based on real-time system constraints (\eg, memory bandwidth);
(3) Formal verification~\cite{hasan2015formal}, which ensures correctness and mitigates fusion-induced computational errors through SMT-based equivalence and boundary checks.

}

\vspace{-1mm}
\add{
\subsection{Future Work}
\label{sec:future}
Current synthesis strategy in \tool{} utilized random pattern insertion. 
However, advanced search strategies could enhance the effectiveness of computational graph synthesis. 
For example, reinforcement learning can be used to guide test generation, where an agent learns to select synthesis points by maximizing the discovery of novel optimization paths in the DL compiler. 
The state space encodes the current coverage of the optimization path, actions modify the seed graph’s synthesis points, and rewards serve as signals to trigger new optimization paths.
Integrating such strategies in \tool{} is a promising future work.
}

\section{Related Work}
\label{sec:related}

\subsection{DL Compiler Testing}
Recently, several techniques have been proposed to test DL compilers~\cite{opera,ma2022hirfuzz,tzer,TVMFuzz,MT_DLComp,whitefox}.
Based on their test generation strategy, they can be primarily categorized into two types: generation-based
and mutation-based techniques.
The former involves NNSmith~\cite{NNSmith}, HirGen~\cite{ma2022hirfuzz}, OPERA~\cite{opera}, TVMFuzz~\cite{TVMFuzz} and WhiteFox~\cite{whitefox}, which utilize predefined grammar or LLMs to generate tests from scratch.
NNSmith generates computational graphs in the DL model format, emphasizing the diversity of operations and structure. 
HirGen produces computational graphs in IR format, incorporating diverse data types and shapes to improve diversity.
OPERA focuses on generating DL models with a single operator for testing the model loading stage.
TVMFuzz focuses on generating hardware-dependent IRs, targeting operator optimizations.
WhiteFox, the state-of-the-art technique, leverages LLMs to mine optimization patterns from compiler source code and generate tests.
The latter includes MT-DLComp~\cite{MT_DLComp} and Tzer~\cite{tzer}, which generate tests by modifying existing ones. 
MT-DLComp~\cite{MT_DLComp} applies semantics-preserving mutations 
to generate equivalent computational graphs for metamorphic testing.
Tzer~\cite{tzer} applies joint mutations on hardware-specific IRs and \textit{passes} to explore diverse program states.

Unlike existing testing techniques, we are the first to introduce the computational graph synthesis technique, called \tool{}, which addresses the challenge of generating optimization-aware computational graphs for testing DL compilers. 
\tool{} identifies optimization-aware patterns from documented tests via computational graph analysis and then designs two synthesis strategies to combine patterns with various contexts, producing valid and diverse optimization-aware computational graphs.

\subsection{Synthesis-Based Compiler Testing}
In traditional compiler testing, synthesis-based techniques~\cite{javatailer,langfuzz,creal} have been demonstrated effective in detecting bugs.
JavaTailor~\cite{javatailer} extracts bug-revealing ingredients from historical bug-triggering tests and combines them into seed programs to generate new tests. 
LangFuzz~\cite{langfuzz} leverages bug-revealing code fragments from bug reports to fuzz JavaScript interpreters. Creal~\cite{creal} extracts C/C++ functions and incorporates them into seed programs by introducing function calls. 

Unlike these techniques, \tool{} aims at synthesizing optimization-aware computational graphs tailored to DL compilers, with distinct test objectives (\ie, DL compiler optimizations) and test formats (\ie, computational graph). 
\tool{} generates optimization-aware computational graphs by extracting patterns from documented tests and tailors synthesis strategies to insert the pattern into diverse contexts. 
Rather than reusing bug-triggering elements from historical tests, \tool{} emphasizes optimization-aware patterns extraction and synthesis for uncovering optimization bugs in DL compilers.
Meanwhile, the test generated by \tool{} is a computational graph, which fundamentally differs from the test for traditional compilers like JVM~\cite{gao2023vectorizing,zhao2024program} and Solidity~\cite{ma2024towards}.
This distinction creates unique technical challenges for \tool{} in generating valid optimization-aware computational graphs, addressed by two novel synthesis strategies: reusing inputs/outputs of existing nodes and creating new connection nodes with compatible inputs/outputs.

\section{Conclusion}
\label{conclusion}
This paper proposes \tool{}, a novel optimization-aware technique for computational graph synthesis. 
First, it extracts optimization-aware patterns from documented tests through code instrumentation.
Then, it synthesizes tests by combining extracted patterns with diverse contexts.
To ensure the validity and optimization-awareness of synthesized computational graphs, two synthesis strategies are designed, including a strategy to reuse the compatible inputs/outputs of the existing nodes in the context and a strategy to create new nodes with compatible inputs/outputs for connections.
Leveraging the implicit test oracles based on crashes and inconsistencies, \tool{} applies the synthesized computational graphs to testing two popular DL compilers, TVM and ONNXRuntime.
The results show that \tool{} significantly outperforms the state-of-the-art techniques in terms of both bug detection and coverage metrics. By fuzzing the latest versions of TVM and ONNXRuntime for two weeks, \tool{} identified \TotalBugsNum{} previously unknown bugs, of which \ConfirmedBugsNum{}/\FixedBugsNum{} have been confirmed/fixed by developers.

\begin{acks}
This work is supported by the National Key Research and Development Program of China (No. 2024YFE0204200), National Natural Science Foundation of China (Grant Nos. 62322208 and 12411530122), and the Hong Kong Research Grant Council/General Research Fund (No. 16206524).
Part of this work was done when Qingchao Shen was a visiting student at HKUST under Dr. Yongqiang Tian.
\end{acks}

\balance
\bibliographystyle{ACM-Reference-Format}
\bibliography{ref}

\end{document}